%% file: main.tex
\documentclass[journal]{IEEEtran}

\ifCLASSINFOpdf
\else
\fi

\usepackage[linesnumbered,ruled,vlined]{algorithm2e} 
\usepackage{subcaption}
\usepackage{import}
\usepackage{multirow}
\usepackage{graphicx}
\usepackage{xcolor}
\usepackage{subcaption}
\usepackage{balance}
\usepackage{epsfig,endnotes}
\usepackage{import}
\usepackage{url}           
\usepackage{commath}
\usepackage{booktabs}       
\usepackage{amsfonts}      
\usepackage{nicefrac}       
\usepackage{microtype}     
\usepackage{array}
\usepackage{stfloats}
\usepackage{epigraph}
\usepackage{hyperref}
\usepackage{soul}
\usepackage{cite}
\usepackage{framed}
\input{macros}
\newtheorem{theorem}{Theorem}

\begin{document}

\title{\name: Federated Learning as a Medium for Covert Communication}

\author{
    \IEEEauthorblockN{
        Dorjan Hitaj\IEEEauthorrefmark{1},
        Giulio Pagnotta\IEEEauthorrefmark{1},
        Briland Hitaj\IEEEauthorrefmark{2}, 
        Fernando Perez-Cruz\IEEEauthorrefmark{3} and
        Luigi V. Mancini\IEEEauthorrefmark{1}\\
    }
    \IEEEauthorblockA{
        \IEEEauthorrefmark{1}Department of Computer Science, Sapienza University of Rome, 
        \{hitaj.d, pagnotta, mancini\}@di.uniroma1.it\\
    }
    \IEEEauthorblockA{
        \IEEEauthorrefmark{2}Computer Science Laboratory, SRI International, 
        briland.hitaj@sri.com\\
    }
    \IEEEauthorblockA{
        \IEEEauthorrefmark{3}Swiss Data Science Center and Computer Science Department ETH Z\"{u}rich,
        fernando.perezcruz@sdsc.ethz.ch
    }
}

\maketitle

\thispagestyle{plain}
\pagestyle{plain}

\import{\sectiondir}{01-abstract.tex}

\IEEEpeerreviewmaketitle

\import{\sectiondir}{02-introduction.tex}
\import{\sectiondir}{03-background.tex}
\import{\sectiondir}{04-threat_model.tex}
\import{\sectiondir}{05-CMDA.tex}
\import{\sectiondir}{06-exp_setup.tex}
\import{\sectiondir}{07-evaluation.tex}
\import{\sectiondir}{08-discussion.tex}
\import{\sectiondir}{09-related_work.tex}
\import{\sectiondir}{10-conclusions.tex}

\bibliographystyle{IEEEtranS}
\bibliography{main}

\end{document}

%% file: macros.tex
\SetKwInput{KwInput}{Input}                
\SetKwInput{KwOutput}{Output}              

\def\name{FedComm\xspace}

%% file: sections/01-abstract.tex
\begin{abstract}
Proposed as a solution to mitigate the privacy implications related to the adoption of deep learning, Federated Learning (FL) enables large numbers of participants to successfully train deep neural networks without revealing the \emph{actual} private training data. To date, a substantial amount of research has investigated the security and privacy properties of FL, resulting in a plethora of innovative attack and defense strategies.

This paper thoroughly investigates the communication capabilities of an FL scheme. In particular, we show that a party involved in the FL learning process can use FL as a covert communication medium to send an arbitrary message. 
We introduce \name, a novel  covert-communication technique that enables robust sharing and transfer of targeted payloads within the FL framework. Our extensive theoretical and empirical evaluations show that \name provides a stealthy communication channel, with minimal disruptions to the training process. 

Our experiments show that \name successfully delivers 100\% of a payload in the order of kilobits before the FL procedure converges.
Our evaluation also shows that \name is independent of the application domain and the neural network architecture used by the underlying FL scheme.

\end{abstract}

%% file: sections/02-introduction.tex
\section{Introduction}
\label{sec:introduction}

\epigraph{The single biggest problem in communication is the illusion that it has taken place.}{George Bernard Shaw}

Deep Learning (DL) is the key factor for an increased interest in research and development in the area of Artificial Intelligence (AI), resulting in a surge of Machine Learning (ML) based applications that are reshaping entire fields and seedling new ones. Variations of Deep Neural Networks (DNN), the algorithms residing at the core of DL, have successfully been implemented in a plethora of domains, including here but not limited to image classification~\cite{Simonyan14verydeep, He2016DeepRL}, natural language processing~\cite{nlp1, nlp2}, speech recognition~\cite{speech2, speech3}, data (image, text, audio) generation~\cite{menick2018generating, passGAN, styleGAN, 9833616}, cyber-security~\cite{de_gaspari_evading_2022, hitaj2023minerva, de_gaspari_reliable_2022, 9257172, pagnotta2023dolos}, and even aiding with the COVID-19 pandemic~\cite{valenciacovidwired2021}.
DNNs can ingest large quantities of training data and autonomously extract and learn relevant features while constantly improving in a given task. However, DNN models require significant amounts of information-rich data and demand hardware to support these models' computation needs. These requirements limit the use of DNNs to institutions that can satisfy these requirements and push entities that do not have the necessary resources to pool their data into third-party resources. 
Strategies like transfer learning alleviate these drawbacks, but their adoption is not always possible. 
Also, as highlighted and emphasized by prior research~\cite{shokri2015privacy, hitaj2017deep}, sharing the data in third-party resources is not a viable solution for certain entities because they would risk potential privacy violations, infringing on current laws designed to protect the privacy and ensure data security.
To address the issues described above, Shokri and Shmatikov~\cite{shokri2015privacy} introduce \emph{collaborative learning}, a DL scheme that allows multiple participants to train a DNN without needing to share their proprietary training data.
In each epoch of the collaborative learning scheme, participants train replicas of the target DNN model on their local private data and share the model's updated parameters with the other participants via a global parameter server. This process allows the participants to train a DNN without having access to other participants' data, or pooling their data in third-party resources.
In the same line of thought, McMahan~et al. propose \emph{federated learning} (FL), a decentralized learning scheme that scales the capabilities of the collaborative learning scheme to thousands of devices~\cite{mcmahan2017communication}, successfully being incorporated into the Android Gboard~\cite{mcmahangboard2017}. Additionally, FL introduces the concept of secure aggregation, which includes an additional layer of security in the process~\cite{secureAggregation, mcmahan2017communication}. Currently, a growing body of work proposes variations of FL-schemes~\cite{triastcyn2019federated, mo2021ppfl, luping2019cmfl}, novel attacks on existing schemes~\cite{hitaj2017deep, melis2019exploiting, bagdasaryan2020backdoor, wang2020attack}, and approaches for mitigating such adversarial threats~\cite{DziedzicBeyondFederation21}.

In this paper, our investigation focuses on the extent to which FL schemes can be exploited.
To this end, we ask the following question:
\textbf{Is it possible for a subset of participants to use the shared model's parameters as a medium to establish a covert-communication channel to other members of an FL training process?}

\begin{figure}[t]
    \centering
    \includegraphics[width=0.85\columnwidth]{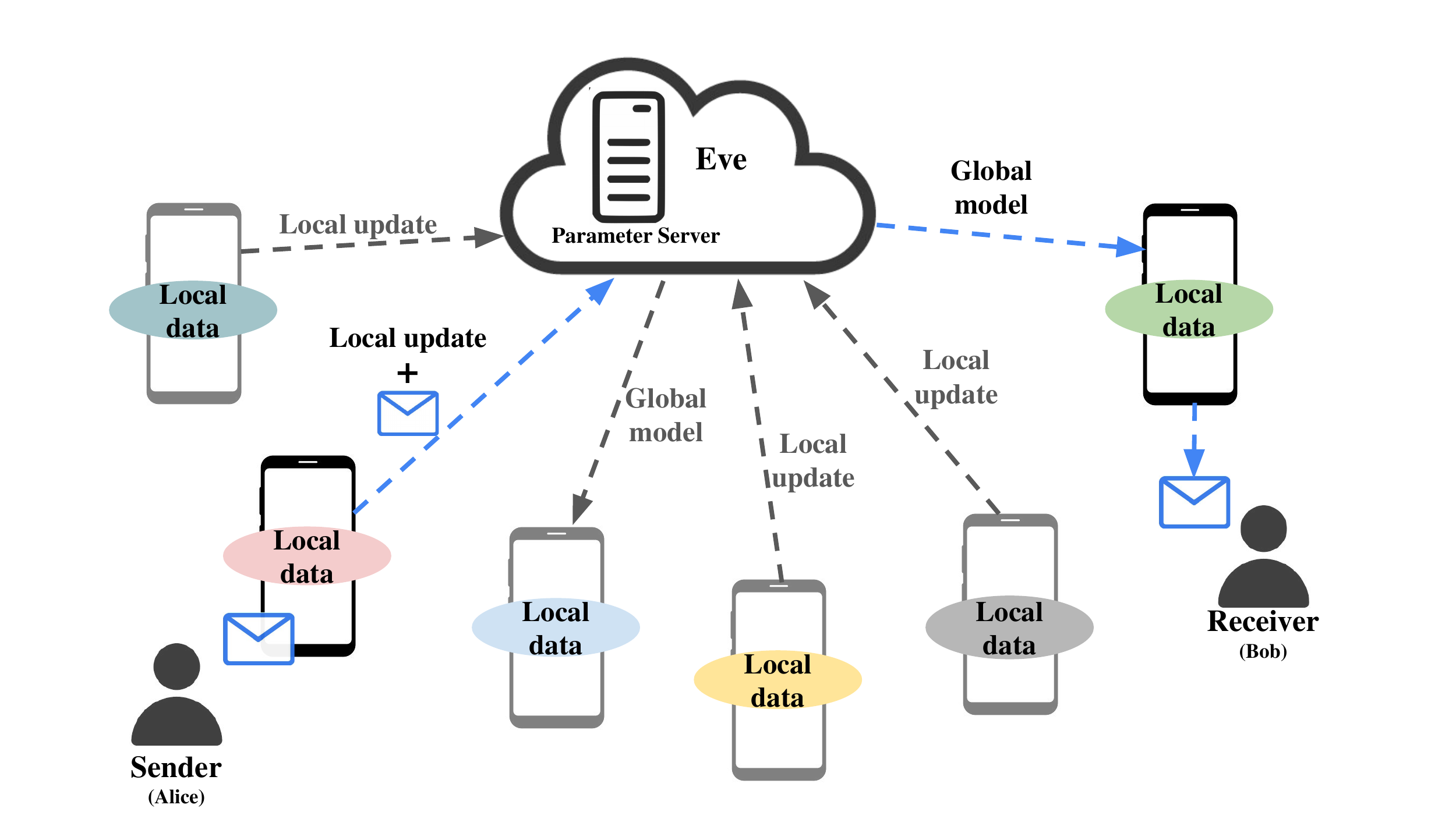}
    \caption{High level overview of the \name communication scheme.}
    \label{fig:fl_scheme}
\end{figure}

We demonstrate that the transmission of information ``hidden'' within the model's parameters is feasible by encoding such information using a combination of Code-Division Multiple Access (CDMA) and LDPC error correction techniques. In particular, CDMA spread-spectrum channel-coding enables covert communication and allows the desired level of ‘stealthiness’ to be adjusted according to the specific application, and we use LDPC to ensure robustness against channel errors arising from  parameter alteration, e.g., noise caused by aggregation of other participants' parameters made by the parameter server.
Our proposed covert-communication scheme, \name, stealthily transmits information that is hidden, e.g. from the global parameter server and from the other participants who are not \emph{senders} or \emph{receivers} of the communication.
Figure~\ref{fig:fl_scheme} shows the high-level operation of the \name scheme. 
First, the \emph{sender} encodes the \emph{message} in its model's updated parameters and then sends them to the global parameter server.
The \emph{receiver} obtains the global model from the parameter server and can precisely decode the message that was sent.
We demonstrate the transmission of covert content varying from simple text messages, e.g., ``\textsc{Hello World!}'', to complex files, such as images. 
Furthermore, we prove theoretically and empirically that the \name scheme causes no disruptions to the FL process and has a negligible effect on the performance of the learned model.
This paper assumes an abstract attacker capable of inspecting and performing statistical analysis on participant parameter updates. An example of such an attacker could be a parameter server maliciously modified for the purpose of monitoring participant parameter updates. Another example of using \name is to support covert communication for citizens or journalists in countries with oppressive regimes. In this scenario, our abstract attacker models the actions of the oppressive regime that has complete control over the nation's infrastructure and is seeking to detect any covered communication channels between participants in an FL scheme.

Our proposed approach does \textbf{not} aim at changing the behavior of the DNN model in the presence of triggers. \name uses an FL scheme as a \textbf{medium} to covertly transmit additional content to other participants without altering the behavior of the resulting ML model.

Our contributions include the following:
\begin{enumerate}
    \item We introduce \name, a novel covert-channel communication technique that can be incorporated seamlessly into existing FL schemes. \name permits entities participating in the FL process to covertly transmit payloads of varying size to other participants.
    
    \item We incorporate Code-Division Multiple Access (CDMA), a spread-spectrum channel-coding technique~\cite{spread_spectrum_principles} that is designed to ensure secure and stealthy military communications and build the foundations of \name around it.

    \item We theoretically demonstrate the feasibility and effectiveness of the proposed approach. 

    \item We demonstrate that \name is domain-independent and we conduct an extensive empirical evaluation under varying conditions: a) diverse payload sizes; b) different DNN architectures; c) multiple domains, including image, text, and audio; and, d) different number of participants selected for updating the model on each FL round.
    
\end{enumerate}
 

%% file: sections/03-background.tex
\section{Background}\label{sec:background}
\subsection{Deep Learning}
\label{sec:deep_learning}

Deep Learning relies heavily on the use of neural networks, which are ML algorithms inspired by the human brain and are designed to resemble the interactions amongst neurons~\cite{machinelearningMitchell}. 
While standard ML algorithms require the presence of handcrafted features to operate, NNs determine relevant features \emph{on their own}, learning them directly from the input data during the training process~\cite{Goodfellow_deeplearning_book}.
Two main requirements underline the success of NNs in general:
1) substantial quantities of rich training data, and 2) powerful computational resources. Large amounts of diverse training data enable NNs to learn features suitable for the task at hand, while simultaneously preventing them from memorizing (i.e., overfitting) the training data. Such features are better learned when NNs have multiple layers, thus the deep neural networks (DNN). 
Research has shown that the single-layer, shallow counterparts are not good at learning meaningful features and are often outperformed by other ML algorithms~\cite{Goodfellow_deeplearning_book}. DNN training translates to vast numbers of computations requiring powerful resources, with graphical processing units (GPUs) a prime example.

\subsection{Federated Learning}
\label{sec:federated_learning}

Federated learning (FL) enables multiple parties to jointly contribute to the training of a DNN without the need to share the actual training data~\cite{mcmahan2017communication, secureAggregation}.
FL is organized in rounds (also referred as epochs), during which the participants train the local DNN over their respective local datasets, and then updated the updated gradients to the parameter server, which, in turn, updates the global model. In doing so, the DNN models get trained without actually ``\emph{seeing}'' other participants' data, thus making FL an attractive alternative for entities interested in benefiting from the DL, but do not possess large quantities of data and powerful resources, or possess sensitive data that cannot be easily distributed.

\leavevmode \\ 
\textbf{How does FL work?}
 Lets denote the weight parameters of a DNN by $\mathbf{W}$.
 FL is typically organized in rounds (time-steps $t$ here). At time $t$, a subset of users, $n'\leq n$ out of $n$ participants that have signed up for collaborating is selected for improving the global model. Each user trains their model and computes the new model:
\begin{equation}
\mathbf{W}_{t+1}^k = \mathbf{W}_t + \alpha \nabla \mathbf{W}^k_t
\label{eq:user_train}
\end{equation}

Where $\mathbf{W_t}$ are the DNN weights at time $t$, and  $\nabla \mathbf{W}^k_t$ is the mini-batch gradient for user $k$ at time $t$. 

Participants of the FL learning scheme send to the global parameter server the gradient $\nabla\mathbf{W}^k_t$. 
On receiving the gradients, the server recomputes the model at time $t+1$, as follows:
\begin{equation}
\mathbf{W}_{t+1} = \mathbf{W}_{t} + \frac{\alpha}{n'} \sum_{k=1}^{n'} \nabla \mathbf{W}^k_t
\label{eq:gradient_averaging}
\end{equation}

where $n'$ is the number of updates obtained by the parameter server (i.e., the number of participants taking part in the current training round) and $\alpha$ is a weight factor chosen by the global parameter server.

\subsection{Code-Division Multiple Access (CDMA)}\label{sec:cdma_background}

In digital communications, spread-spectrum techniques~\cite{spread_spectrum_principles}
are methods by which a signal (e.g., an electrical, electromagnetic, or acoustic) with a particular bandwidth is deliberately spread in the frequency domain. These techniques enable the spreading of a narrowband information signal over a wider bandwidth. On receiving the signal, the receiver knowing the spreading mechanism can recover the original bandwidth signal. These techniques were developed in the 1950s for military communications because they resist the enemy's efforts to jam the communication channel and hide the fact that the communication is taking place~\cite{Verdu98}. 
Practically, two main techniques are used to spread the bandwidth of a signal: \textit{frequency hopping} and \textit{direct sequence}. In \textit{frequency hopping}, the narrowband signal is transmitted for a few (milli or micro) seconds in a given band that is constantly changed using a pseudo-random frequency band that has been agreed upon with the receiver. The receiver, in coordination, tunes its filter to the agreed-on frequency band to recover the message. 
\textit{Direct Sequence}, the spreading technique we use in \name, works by directly coding the data at a higher frequency by using pseudo-random generated codes that the receiver knows. 
In the 1990s, Direct Sequence Spread Spectrum was proposed as a multiple-access technique (i.e., Code Division Multiple Access or CDMA) in the IS-95 standard for mobile communications in the US, and it was adopted worldwide as the Universal Mobile Telecommunications System (UMTS) standard in the early 2000s. This standard is better known as 3G.
In CDMA, if several mobile users want to transmit information to a base station, they all transmit at the same time over the same frequency but with different codes.
The base station would correlate the code of each user with its spreading code to detect the transmitted bits. The other users' information would contribute to the noise level.
In CDMA, if we make the spreading code long enough, the transmitted sequence can be hidden under the noise level, making it non detectable by the unintended user, but the receiver can recover it once we add all the contributions from the code.
Typically, the spreading codes are in the hundreds to thousands, so the signal is only visible when the spreading code is known, and the gain of using CDMA is proportional to the length of the spreading code.

\subsection{Covert Communication Channels}\label{sec:covert_channels}
A covert channel~\cite{covert_channels} is an indirect communication channel between unauthorized parties that violates a security policy by using shared resources in a way in which these resources are not initially designed, bypassing mechanisms that do not permit direct communication between these unauthorized parties in the process. As such, covert channels emerge as a threat to information-sensitive systems in which leakage to unauthorized parties may be unacceptable (e.g., military systems).
Initial research in covert channels focused on \emph{single-systems} (i.e., File-lock, disk-arm, and bus-contention covert channels)~\cite{gallager_book,Schafer,karger}. However, shortly after, covert channel techniques encompassing multi-systems, such as a network of devices, were also devised, giving birth to a new kind of threat, that of multi-system covert channels~\cite{characterizing_network_covert_channels}. 
Multiple techniques have been crafted to set-up \emph{multi-system} covert-communication channels ranging through a variety of protocols including TCP, IP, HTTP, ICMP, AODV, and MAC protocols~\cite{4630112,http_covert_channels,covert_channel_ad_hoc,griffin_tcp_timestamps}. The core concept behind these covert-communication approaches is encapsulating the covert channels in legitimate protocols to bypass firewalls and content filters. Multi-system covert channels are an advanced class of security threats to distributed systems. Using this type of covert channels, an adversary can exfiltrate secret information from compromised machines without raising suspicion from firewalls, which typically inspect only the packet payload.
There are two major categories of multi-system covert channels: i) timing-based and ii) storage-based.
Timing-based multi-system covert channels~\cite{IP_covert_timming,Gianvecchio,lampson_confinement,4630112} can exfiltrate secret data by modulating the inter-packet delays (IPDs) of network traffic, e.g., by using large (small) IPDs to encode ones (zeros). 
Storage-based multi-system covert channels exploit optional or unused TCP/IP header fields, such as ToS, Urgent Pointer, and IPID fields~\cite{stego_active_warden}, essential header fields such as the TCP initial sequence number~\cite{7087183} or even relying on the inherent non-determinism of network traffic, e.g., embedding data into the receive window size or ACK fields~\cite{5198826}.
Timing-based multi-system covert channels can be detected and mitigated by observing the network traffic distribution because modulated traffic traces would have different IPD distributions from regular network traffic. Performing statistical analysis on the network traffic and looking for statistical deviations between a given IPD distribution and a known-good distribution of regular network traffic can make it possible for the detection of an ongoing covert communication channel~\cite{1624024}. After detection, these timing-based channels can be mitigated by buffering or injecting random delays to network traffic in order to disrupt the IPD modulation~\cite{7870221}.
Storage-based multi-system covert channels can be detected by inspecting all header fields and looking for the existence of header fields that are rarely used or contain suspicious values~\cite{7087183,netwarden}. 
However, the giant leaps the technological advancements of the last decade have broadened the space of opportunities for the adversaries. In this paper, we move away from ``traditional'' mediums for multi-system cover channels and shed light to alternatives that make use of new, yet unconventional platforms, for covert communication, such as federated learning schemes.

%% file: sections/04-threat_model.tex
\section{Threat Model and System Overview}
\label{sec:threat_model}
Our threat model is composed of three actors: Alice, Eve, and Bob, see Figure~\ref{fig:fl_scheme}. Alice represents the \emph{sender}, a participant in the FL scheme aiming to secretly communicate information to Bob (the \emph{receiver} participant), while Eve is an abstract attacker capable of inspecting and performing statistical analysis on participant parameter updates. An example of Eve could be a parameter server modified for the purpose of detecting any covered communication channels among participants.
To achieve her goal, Alice needs to exploit a shared resource, setting up a \emph{covert communication channel} with Bob. Alice succeeds if she can secretly share information with Bob, while remaining undetected by Eve. 
\leavevmode \\
\textbf{Actors' Objectives and Capabilities.}
\begin{itemize}
    \item Alice and Bob are two (or more) participants in an FL scheme interested in covertly communicating with one another.
    \item Alice's objective consists in conveying (transmitting) a \emph{secret} message to Bob.
    \item Alice makes use of the FL platform as a medium for communicating with Bob. This is a useful strategy, particularly in scenarios where ``traditional'' covert communication channels fail or are impossible to deploy due to firewalls or intrusion detection systems, Section ~\ref{sec:covert_channels}. 
    \item Differently from prior work~\cite{hitaj2017deep, melis2019exploiting}, Alice and Bob do not seek to violate the privacy of the other participants in the FL scheme.
    \item Furthermore, Alice and Bob do not control the global parameter server nor seek to compromise its operation.
    \item Similar to other participants, Alice has a local replica of the DNN model that needs to be trained via FL.
\end{itemize}
\vspace{-1em}
\leavevmode \\
\textbf{\name Communication Overview.}
\begin{itemize}
    \item Alice and Bob make use of \name, our proposed technique, to establish a covert communication channel on top of FL.
    \item Alice and Bob agree on the settings of the encode-decode procedure beforehand.
    \item Using \name, Alice embeds and hides a desired message in her model parameters. The updated parameters are then shared with the other participants via the global parameter server.
    \item While multiple participants (if not all) receive the updated model, only Bob is able to decode the hidden message due to the fact that only Bob and Alice know the encode-decode procedure that has to be followed.
\end{itemize}

We bring to the attention of the reader that federated learning relies on secure aggregation~\cite{secureAggregation} where the parameter server is oblivious of the individual updates and does not possess any tracking mechanism. In our threat model, we assume that our attacker (Eve) is more powerful and is able to inspect and perform statistical analysis on the participant's parameter updates. This enables us to thoroughly evaluate the stealthiness of \name's covert communication channel.

\leavevmode \\
\textbf{Possible application in real life.}
Following our threat model, Eve can inspect and perform statistical analysis on participant parameter updates. We envision uses for \name in cases where traditional covert communication systems would fail or be difficult to deploy such as covert communication for citizens or journalists in countries with oppressive regimes. Note that these regimes have complete control over devices, networks and traffic within the country. In this scenario, Eve models the oppressive regime's actions to monitor participant parameter updates. In this situation, \name can be a valuable asset for a journalist (i.e., Alice) to exfiltrate sensitive information and communicate to entities external (i.e., Bob) to the oppressive regime without being detected. For simplicity, in the following sections, we refer to the aggregator (i.e., parameter server) as being Eve.

%% file: sections/05-CMDA.tex
\section{\name}
\label{sec:fedcomm}

This section introduces \name, our covert-communication channel technique built on top of the FL scheme. In its core, \name employs CDMA to build this covert communication channel. In this view, each weight of the NN is a chip (in CDMA parlance) in which we can encode information. CDMA channel synchronization is achieved in \name by an ordering of the weights of the NN that is predefined by the sender and the receiver\footnote{In our implementation, the sender and receiver share a seed, which they use to generate the ordering of the weights of the NN necessary for the channel synchronization.}.

Let's assume that the \textit{sender} aims to transmit a payload of $P$ bits $\mathbf{b} = [b_0,\ldots,b_{P-1}]$. Each bit is encoded as $\pm1$ and the code for each bit is represented by $\mathbf{c}_i$, which is a vector of $+1$ and $-1$ of the same length of the vector $\mathbf{W}$.
For simplicity, consider the weight parameters $\mathbf{W}$ of the NN as a vector.
We denote by $R$ the number of weight parameters $\mathbf{W}$ of the neural network selected to carry the message. Suppose we have a message consisting of $P$ bits. Then, matrix $\mathbf{C}$ has dimension $R$ by $P$, meaning that each bit of the message has an individual spreading code of length $R$.
$\mathbf{C}$ is known only by the \textit{sender} and the \textit{receiver}. 
We assume that the codes have been randomly generated with equal probabilities for $\pm 1$. 
The \textit{sender} is part of a federated learning scheme where $n$ users have signed up for collaborating.
The parameter server proposes a set of initial weights $\mathbf{W}_0$, which are distributed to all the users. 
For simplicity, we view each FL round, where the participants perform a round of training over their respective local datasets as a time instant of the CDMA communication. At every iteration $t$, each user will use their data or a mini-batch of their data to compute the gradient $\nabla \mathbf{W}^k_t$ for $k=0, \ldots, n-1$.
Without loss of generality, we assume that the \textit{sender} is user 0. The \textit{sender} encodes the payload on its gradient as follows: 

\begin{equation}
    \widehat{\nabla \mathbf{W}^0_t} = \beta\nabla \mathbf{W}^0_t + \gamma \mathbf{C}\mathbf{b} \label{mod_grad}
\end{equation}
 
where $\gamma$ and $\beta$ are two gain factors to control the \textit{power}\footnote{In this paper, \textit{power} refers to the magnitude of the gradient. The magnitude of a gradient is calculated by taking the square root of the sum of each of the individual gradient values squared (i.e., $l2$ norm). We use magnitude and power interchangeably.} of the gradient update of the sender.
We introduce these two tunable parameters because a signal being added using spread-spectrum techniques can be detected by measuring the power of the overall signal in the channel (i.e., the magnitude of the gradient update containing the message). If that power is larger than expected, the presence of an added signal can be detected~\cite{965114}. We choose values of $\gamma$ and $\beta$ in such a way that the power of the modified gradient (i.e., the gradient that contains the message) is like the power of the unmodified gradient. In particular, considering a $P$-bit payload and a gradient update with variance $\sigma^2$ we select $\gamma = \delta \sigma/\sqrt{P}$ where $\delta$ controls the power of the payload signal and thus directly impacts the value of $\beta = \sqrt{1-\delta^2}$ which will change the power of the actual gradient such that the overall update sent to the parameter server has the same power as the gradient update had before embedding the payload in it.
Next, the parameter server updates the weights of the network using Eq~\eqref{eq:gradient_averaging} and distributes $\mathbf{W}_t$ to all of the users.
After $T$ rounds, where $T$ is the amount of transmission rounds needed to deliver the message~\footnote{We show how to estimate $T$ in the second part of the proof of Theorem 1.}, it is possible to recover the payload. The \textit{receiver} can recover the payload that was hidden in the gradients by correlating the weights of the global model $W_T$ with the spreading codes $\mathbf{C}$.

For example, for bit $i$, the receiver can recover the payload as follows:
\begin{align}
    y_i = \mathbf{c}_i^\top \left(\mathbf{W}_T-\mathbf{W}_0\right)\label{eq:decode_y}\\
    \widehat{b}_i = \text{sign}(y_i)\label{eq:decode}
\end{align}

where $\mathbf{c}_i^\top$ is the transpose of the spreading code for bit $i$.
Note that the receiver can recover the payload for any other round $t > T$ too.
In the following, we prove the main properties of \name.
\begin{framed}\noindent
\begin{theorem}
Let ${b}_i$ be the bit that needs to be sent using \name. Under the assumption that in each round the gradient update of each participant behave as zero mean Gaussian with constant variance  $\sigma^2$: 
\begin{enumerate}
    \item The distribution of ${b}_i$ is given by \begin{align} \mathcal{N}\left(b_i, \frac{P-1}{R}+\frac{n\sigma^2}{TR\gamma^2}\right) \end{align}
    \item Given $\gamma = \delta \sigma^2/\sqrt{P}$, to transmit a $P$-bit payload there are needed at least $nP/(\delta^2(R-P))$ global rounds.
\end{enumerate}
\end{theorem}\label{theorem}
\end{framed}

\leavevmode \\
\textbf{1. Distribution of ${b}_i$.}
To prove the first part of Theorem 1, we need to delve deeper into the sources of signal that compose the $y_i$ (Eq.~\ref{eq:decode_y}).  Bit $y_i$ is composed of three signal sources, namely the signal of the i-th bit $b_i$ that we are trying to decode, the noise coming from the gradients of the other participants, and the signal coming from the other bits of the payload. 
Let's see why it is that!
\begin{align}
y_i =& \mathbf{c}_i^\top \left(\mathbf{W}_T-\mathbf{W}_0\right)\\
 = &\frac{\alpha}{n}\mathbf{c}_i^\top \left(\sum_{t=0}^{T-1}\left(\widehat{\nabla \mathbf{W}^0_{t}}+ \sum_{k=1}^{n-1} \nabla \mathbf{W}^k_t  \right)\right)\label{iteration_t_prime} \\
 = & \frac{\alpha}{n} \left(T\gamma \mathbf{c}_i^\top\mathbf{C}\mathbf{b} + \mathbf{c}_i^\top \sum_{t=0}^{T-1}\left(\beta\nabla \mathbf{W}^k_0+\sum_{k=0}^{n-1} \nabla \mathbf{W}^k_t \right)\right)\\
 = & \frac{\alpha}{n} \left(T\gamma \mathbf{c}_i^\top\mathbf{c}_ib_i+T\gamma \mathbf{c}_i^\top\mathbf{C}_{\neg i}\mathbf{b}_{\neg i} + \mathbf{c}_i^\top \widetilde{\mathbf{w}} \right)\label{breakC}\\
 = & \frac{\alpha}{n} \left(T\gamma \mathbf{c}_i^\top\mathbf{c}_ib_i+T\gamma \mathbf{c}_i^\top \widetilde{\mathbf{c}}  + \mathbf{c}_i^\top \widetilde{\mathbf{w}} \right)\label{breakC2}\\
 = &\frac{\alpha}{n}\left(T \gamma R b + \epsilon_i^\mathbf{C} + \epsilon_i^\mathbf{W} \right) =\frac{\alpha}{n}\left(T \gamma R b + \epsilon_i \right)
\end{align}

To derive Eq~\eqref{breakC}, we have divided matrix $\mathbf{C}$ in a column vector $\mathbf{c}_i$ and a matrix $\mathbf{C}_{\neg i}$ that contains all columns except the $i$-th column, resulting in a $R\times P-1$ matrix. The vector $\mathbf{b}_{\neg i}$ is a $P-1$ dimensional vector, which is only missing the $i$-th entry. In Eq~\eqref{breakC}, we have also defined $\widetilde{\mathbf{w}} = \sum_{t=0}^{T-1}\left(\beta\nabla \mathbf{W}^k_0+\sum_{k=0}^{n-1} \nabla \mathbf{W}^k_t \right)$. 
In Eq~\eqref{breakC2}, we have defined $\widetilde{\mathbf{c}}=\mathbf{C}_{\neg i}\mathbf{b}_{\neg i}$. The distribution for each component of $\widetilde{\mathbf{c}}$ is a symmetric binomial distribution between $\pm (P-1)$, because the entries of both $\mathbf{C}_{\neg i}$ and $\mathbf{b}_{\neg i}$ are $\pm 1$ with equal probabilities. When we multiply this vector by $\mathbf{c}_i$ and add all the components together, we get a binomial distribution with values between $\pm R(P-1)$. Hence the distribution for $\epsilon_i^\mathbf{C}$ for large $R$ can be approximated by a zero-mean Gaussian with variance $T^2\gamma^2 R(P-1)$.

To compute the distribution of $\epsilon_i^\mathbf{W}$, we assume for now that $\nabla \mathbf{W}_t^k$ is a zero-mean with a variance $\sigma^2$\footnote{For tractability of the theoretical analysis, we assume that the gradients from all the users in every FL round are equally distributed, as a zero-mean Gaussian with a fixed variance $\sigma^2$. We direct the reader to Section~\ref{sec:message_delivery_time} for the hypothesis validation.}. Each component of $\widetilde{\mathbf{w}}$ adds up $T(n-1+\beta)$ of these values. From the central limit theorem (large enough $T(n-1+\beta)$), each one of these variables would be zero-mean Gaussian with a variance~\footnote{We have simplified $(n-1+\beta)$ to $n$, as we will fix $\beta\leq 1$ and we assume $n$ is large enough. In scenarios where $n$ is not too large, we use an upper bound to the variance of $\epsilon_i^\mathbf{W}$.}
$Tn\sigma^2$. When we multiply this vector by $\mathbf{c}_i$ and add all the components together, we end up with a zero-mean Gaussian with a variance $RTn\sigma^2$, because the components of $\mathbf{c}_i$ are $\pm 1$; therefore, they do not change the distribution of the components of $\widetilde{\mathbf{w}}$.

Finally, given that $\mathbf{C}$, $\mathbf{b}$ and $\nabla \mathbf{W}^k_t$ are mutually independent, the variable $\epsilon_i$ is zero mean with a variance that it is the sum of the variances of $\epsilon_i^\mathbf{C}$ and $\epsilon_i^\mathbf{W}$ and also Gaussian distributed. The distribution of $y_i$ is given by $y_i\sim \mathcal{N}(T\gamma R b_i, T^2\gamma^2R(P-1)+RTn\sigma^2)$~\footnote{We have dropped $\alpha/n$ without loss of generality}, and if we further normalized it by $T \gamma R$, it leads to:
\begin{align}
y_i\sim &\mathcal{N}\left(b_i, \frac{T^2\gamma^2R(P-1)+TRn\sigma^2}{T^2\gamma^2R^2}\right)\\& \mathcal{N}\left(b_i, \frac{P-1}{R}+\frac{n\sigma^2}{TR\gamma^2}\right)\label{eq:variance}
\end{align}

To recover $b_i$, it is necessary for the variance of $y_i$ to be less than one~\cite{Cover06}; note that how this can be ensured by \name will be shown in part 2 of this proof.
In particular, considering the results in~\cite{Cover06}, i.e. with a long enough error-correcting code, errorless communication can be ensured when the variance of $y_i$ is at most 1, in \name we employ LDPC error-correcting codes to make possible the decoding of the received bit correctly, thus $\widehat{b}_i = b_i$.

\leavevmode \\
\textbf{2. Message Delivery time.}
Here we show that transmitting a payload of $P$-bits using \name requires at least $nP/(\delta^2(R-P))$ global rounds. Substituting $\gamma = \delta \sigma/\sqrt{P}$ in Eq.\eqref{eq:variance} the variance of $b_i$ becomes: 
\begin{align}
\frac{P-1}{R}+\frac{n\sigma^2}{TR \gamma^2} &= \frac{P-1}{R}+\frac{n\sigma^2}{TR \left(\frac{\delta\sigma}{\sqrt{P}}\right)^2}\label{eq:varcalc}
\\
&= \frac{P-1}{R}+\frac{\delta^2nP}{TR} \approx \frac{(T+\delta^2n)P}{TR}
\end{align}

As mentioned above, this variance must be less than 1 to be able to correctly decode the transmitted bit $b_i$. 
Given this condition and the chosen value of $\gamma = \delta \sigma/\sqrt{P}$, the number of iterations that we need before the message can be seen by the receiver is $T>nP/(\delta^2(R-P))$.

\begin{framed}\noindent
This concludes the proof of Theorem 1. Specifically, we showed that:

\begin{enumerate}
    \item The distribution of ${b}_i$ is given by \begin{align} \mathcal{N}\left(b_i, \frac{P-1}{R}+\frac{n\sigma^2}{TR\gamma^2}\right) \end{align}
    \item Given $\gamma = \delta \sigma^2/\sqrt{P}$, to transmit a $P$-bit payload there are needed at least $nP/(\delta^2(R-P))$ global rounds.
\end{enumerate}

In Section~\ref{sec:evaluation}, we test these theoretical findings in practical applications of FL scenarios.
\end{framed}

\subsection{Remarks} 
\leavevmode \\
\textbf{Message delivery time.}
If we need to add the payload faster, we can add the same payload by tallying more users that transmit the same information with the same code. This information would be added coherently, even if the users are not transmitting the information at the same time, and would lead to an amplification of the message without additional noise. If we have $M$ \textit{senders}, instead of one, adding the same payload with the same code to their gradients, $y_i$ distribution would be equal to $\mathcal{N}(MT\gamma R b_i, M^2T^2\gamma^2R(P-1)+RTn\sigma^2)$. In this case, the payload will be visible $M^2$ times quicker, i.e. $T>nP/(M^2\delta^2(R-P))$.
In the derivation above, we have assumed that all the users send their gradient in each iteration and that all the gradients are used to update the weights. If only a subset $n'<n$ of users are included in each iteration, the analysis above would hold if we define each iteration as being $n/n'$ communication rounds. If the parameter server uses a round-robin scheme, the analysis will be exactly the same. If the parameter server chooses the participants' gradients at random, the message delivery time would hold in mean and, given that the number of rounds in FL is usually large, the deviation in rounds would be negligible (We also evaluate this case in Section~\ref{sec:evaluation}).
As a final note, if we do not have access to $\mathbf{W}_0$ when doing the decoding, we would have an additional error source in $\epsilon_i$, coming from the initialization of the weights $\mathbf{c}_i^\top\mathbf{W}_0$. This would become negligible as $T$ and $n$ grows.

\leavevmode \\
\textbf{Employed Error Correcting codes.}
We employ the Low-Density Parity-Check codes to correct possible errors and thus retrieve $b_i$ correctly. 
In general, Shannon coding theorem~\cite{Cover06} tells us the limit on the number of errors that can be corrected for a given redundancy level, as the number of bits tends to infinity. Low-Density Parity-Check (LDPC) codes~\cite{Richardson08} are linear codes that allow for linear-time decoding of the received word, quasi-linear encoding, and approach the capacity as the number of bits tend to infinity. Linear codes are defined by a coding matrix $\mathbf{G}$. The matrix is a $k\times P$ matrix that transforms $k$ input bits into a sequence of $P$ bits, i.e.
\begin{equation}
\mathbf{b} =  (\mathbf{m}\mathbf{G})\mod 2
\end{equation}
All of the operations are binary operations. In general, linear codes can be described over any Galois field. For simplicity, we will only consider binary codes. 

The bits in $\mathbf{b}$ are then transmitted through an additive noise communication channel: 
\begin{equation}\label{channel_mod}
\mathbf{r} =  (\mathbf{b}+\mathbf{e})\mod 2
\end{equation}

At the receiver, we can check if the received word is valid using the parity-check matrix: 
\begin{equation}
\mathbf{s}=(\mathbf{r}\mathbf{H}^\top)\mod 2.
\end{equation}
The parity-check matrix $\mathbf{H}$ has $P-k$ rows and $n$ columns and spans the null space of complement of $\mathbf{G}$. $\mathbf{s}$ is known as the syndrome and describes the error in the received word. The syndrome is independent of the code-word. If $\mathbf{s} = \mathbf{0}$, the received word is a correct code-word. 

LPDC codes rely on parity check matrices with a vanishing number of ones per column as the number of bits grows. These codes can be proven to approach capacity as the number of bits increases and have an approximate decoding algorithm, i.e., Belief Propagation, that runs in linear time (see~\cite{Richardson08} for further details). The decoding algorithm can also be applied when the channel in equation~\eqref{channel_mod} is not binary. For example, $\mathbf{e}$ can be a Gaussian random variable.
The Belief propagation decoder needs to know the distribution of the errors. If $\mathbf{e}$ is a Bernoulli distributed, we need to know the probability of taking the value 1 and flipping a bit. If $\mathbf{e}$ is Gaussian distributed we need to know its variance.

For our implementation, we rely on a rate-1/2 ($k=p/2$) codes with three ones per row of $\mathbf{H}$.
Once the message has been encoded, we append 100 bits to form the transmitted code-word.  Those 100 bits will be used to estimate the noise level and decode the received bit. These 100 bits are randomly generated and shared between the transmitter and receiver.

%% file: sections/06-exp_setup.tex
\section{Experimental Setup}\label{sec:exp_setup}

We conduct a thorough and extensive evaluation of our proposed scheme considering: 1) a range of benchmark image~\cite{mnist_dataset, krizhevsky2009learning}, text~\cite{wikitext_dataset}, and audio~\cite{piczak2015dataset} datasets; 2) well-known convolutional neural network (CNN), and recurrent neural network (RNN) architectures~\cite{cnn_citation, lstm_citation}; and 3) different payload sizes. This evaluation demonstrates that \name is domain- and model-independent and can be generalized for future areas where FL is deployed.

\subsection{Datasets}
We used the following datasets in our experiments.

The \underline{MNIST} handwritten digits dataset consists of 60,000 training and 10,000 testing grayscale images of dimensions 28x28-pixels, equally divided among 10 classes (0-9)~\cite{mnist_dataset}.

The \noindent\underline{CIFAR-10} dataset is another benchmark image dataset consisting of 50,000 training and 10,000 testing samples of 32x32 colour images divided in 10 classes, with roughly 6,000 images per class~\cite{krizhevsky2009learning}. 

The \noindent\underline{WikiText-2} language modeling dataset, a subset of the larger WikiText dataset, which is composed of approximately 2.5 million tokens representing 720 Wikipedia articles divided into 2,088,628 train tokens, 217,646 validation tokens, and 245,569 testing tokes~\cite{wikitext_dataset}.

The \noindent\underline{ESC-50} dataset consists of 2,000 labeled environmental recordings equally balanced between 50 classes of 40 clips per class~\cite{piczak2015dataset}.

In our evaluation, we partition the datasets in a non independent and identically distributed (non-iid) fashion to emulate the real life deployments of federated learning. 

\subsection{DNN Architectures}\label{sec:dnn_arch_exp}
We adopted different DNN models depending on the task. For the image classification tasks on MNIST and CIFAR-10, we used two CNN-based architectures: a) a standard CNN model composed of two convolutional layers and two fully connected layers; b) a modified VGG model~\cite{Simonyan14verydeep}.
To address the text classification tasks on WikiText-2, we used an RNN model composed of two LSTM layers. 
For the audio classification, we used a CNN model composed of four convolutional layers and one fully connected layer.

\subsection{Transmitted Messages}
We used three different payloads of different sizes for transmission in our covert communication approach for federated learning. The smallest payloads were two text messages of 96 and 136 bits corresponding to the \textit{hello world!} and \textit{The answer is 42!} text phrases. The third payload is a 7904 bits image. For simplicity, we refer to the text messages as SHORT, and the image as the LONG message. 

%% file: sections/07-evaluation.tex
\section{\name Evaluation}\label{sec:evaluation}

This section focuses on the evaluation of \name. We rigorously assessed the effectiveness of the proposed scheme along three main axes:
i) stealthiness, ii) impact on the overall model performance, and iii) message delivery time (the total number of global rounds needed for the receiver to detect the presence of the message).
The following sections provide a step-by-step analysis of all of these metrics.

\begin{figure*}[htp]
	    \centering
	     \begin{subfigure}{.49\textwidth}
            \centering
            \includegraphics[width=.8\columnwidth]{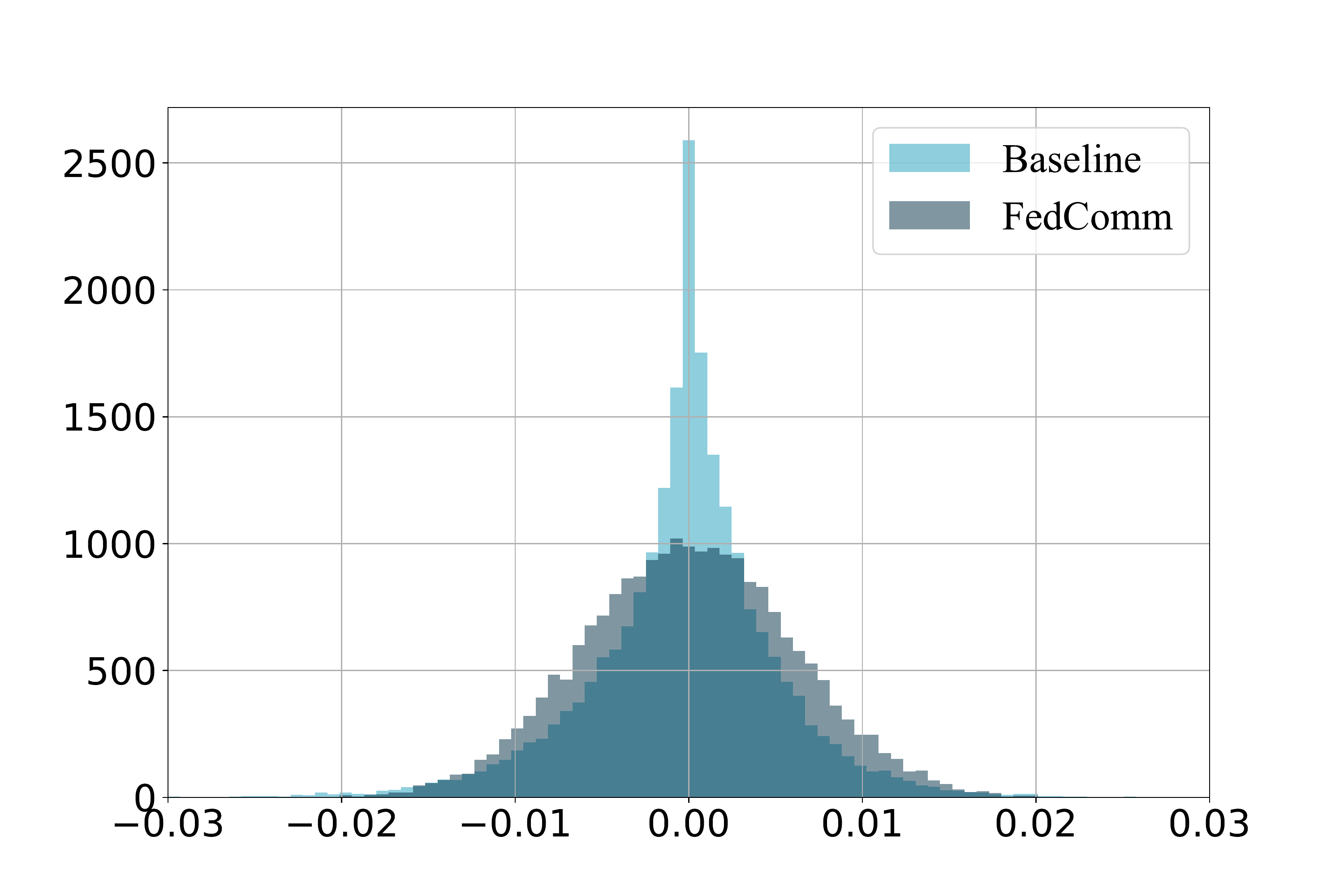}
            \caption{Distribution of gradient update when \name is running non-stealthy with parameters $\beta$=0, $\gamma = \sigma/\sqrt{P}$ vs distribution of gradient update when \name is not running.}    
             \label{fig:non_stealthy_histogram}
        \end{subfigure}
        \hfill
	    \begin{subfigure}{.49\textwidth}
	        \centering
            \includegraphics[width=.8\columnwidth]{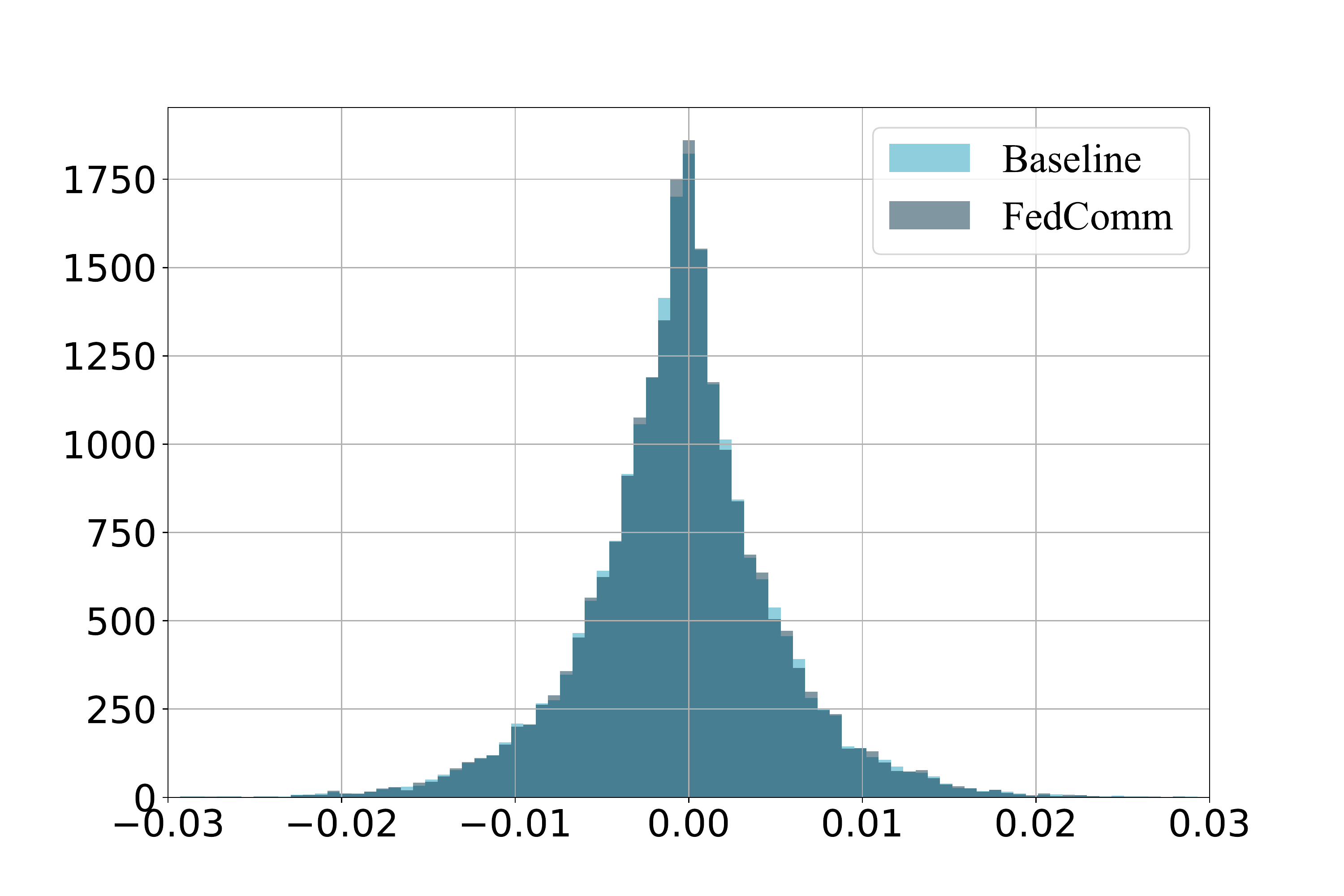}
            \caption{Distribution of gradient update when \name is running full-stealthy with parameters $\beta$=0.99, $\gamma = 0.1\sigma/\sqrt{P}$ vs distribution of gradient update when \name is not running.}
            \label{fig:full_stealthy_histogram}
        \end{subfigure}
        \caption{Stealthiness level; A comparison of the distribution of the gradients when \name is running on different stealthiness parameters and the gradients when \name is not running (i.e. baseline). The source of the data used to obtain the plots are the sender's gradient updates (with and without the embedded message) in a round, and correspond to $\sim$22K datapoints, which is the number of parameters of one of the neural network architectures presented in Section~\ref{sec:dnn_arch_exp})}
        \label{fig:distribution_comparison}
\end{figure*}

\begin{figure}[h]
    \centering
    \includegraphics[width=.8\columnwidth]{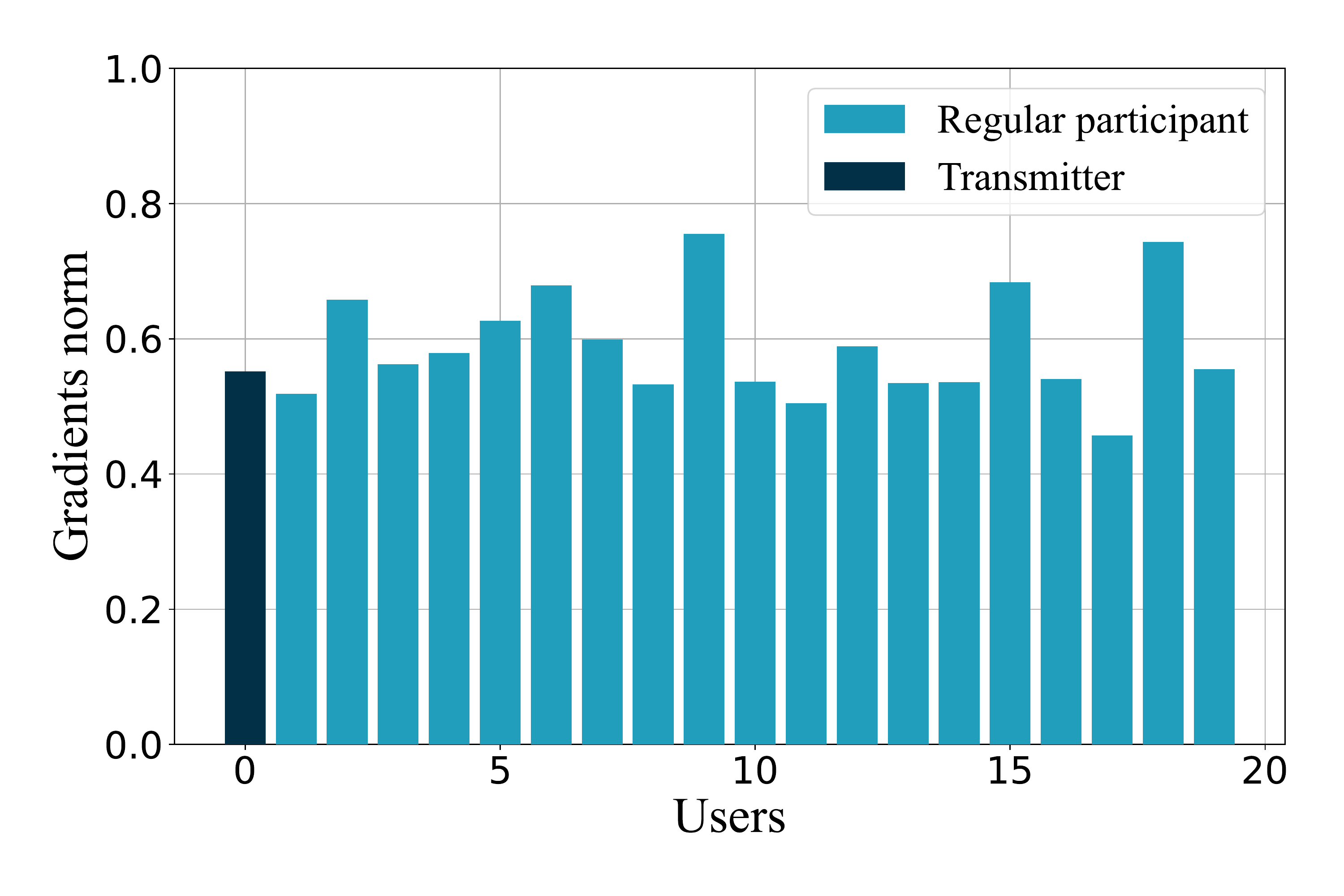}
    \caption{Gradient update norm comparison in a round between regular participants and the \textit{sender} that employs \name with stealthiness parameters $\beta$=0.99, $\gamma = 0.1\sigma/\sqrt{P}$.}
    \label{fig:norm_injection_round}
\end{figure}

\begin{figure*}[htp]
    \centering        
	    \begin{subfigure}{.3\textwidth}
            \centering
            \includegraphics[width=\columnwidth]{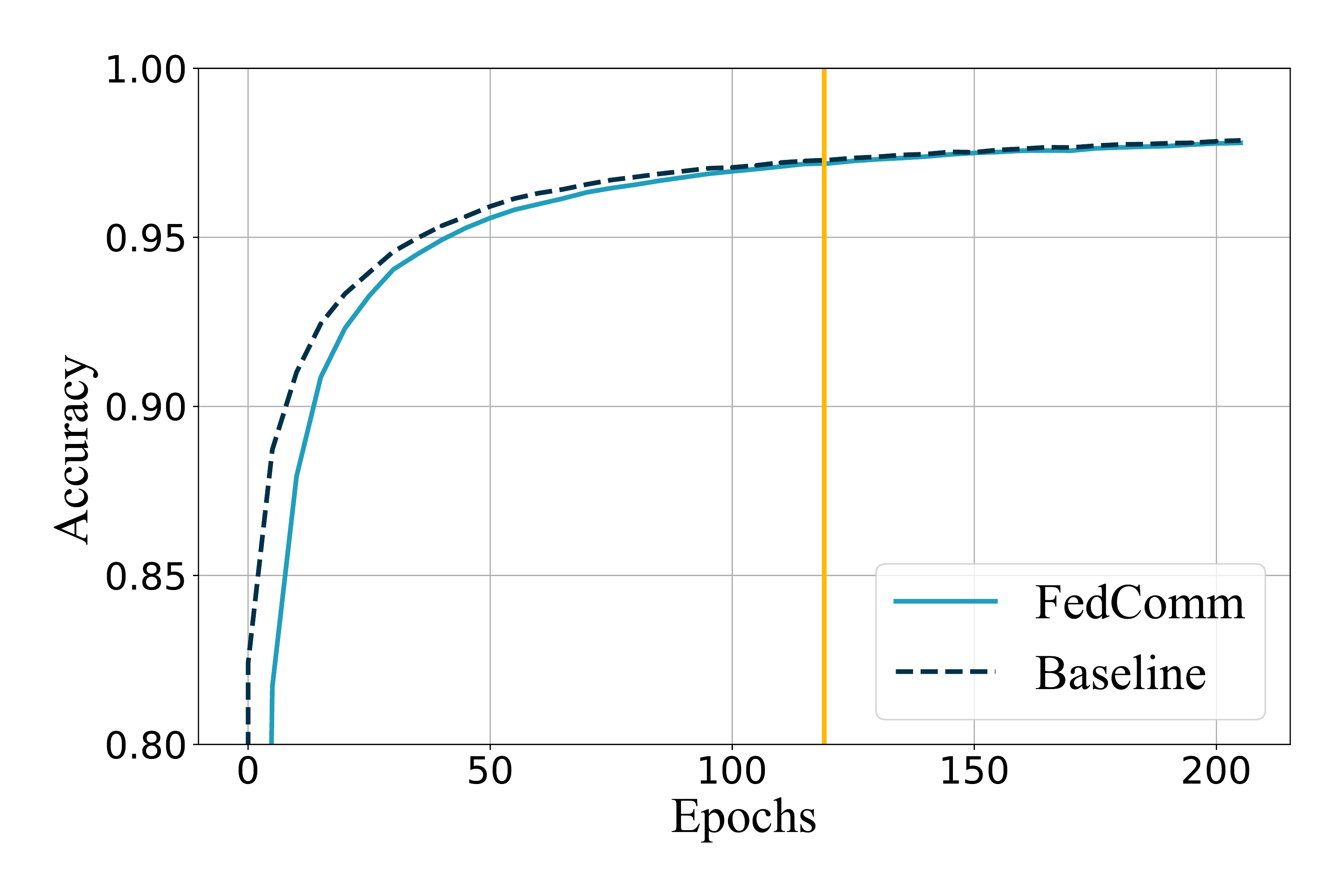}
             \caption{$\beta=0$, $\gamma = \sigma/\sqrt{P}$ \\ 1 sender, short message, MNIST\\ 100\% aggregated per round}
             \label{fig:one_non_stealthy_mnist}
        \end{subfigure}
        \hfill
	    \begin{subfigure}{.3\textwidth}
            \centering
            \includegraphics[width=\columnwidth]{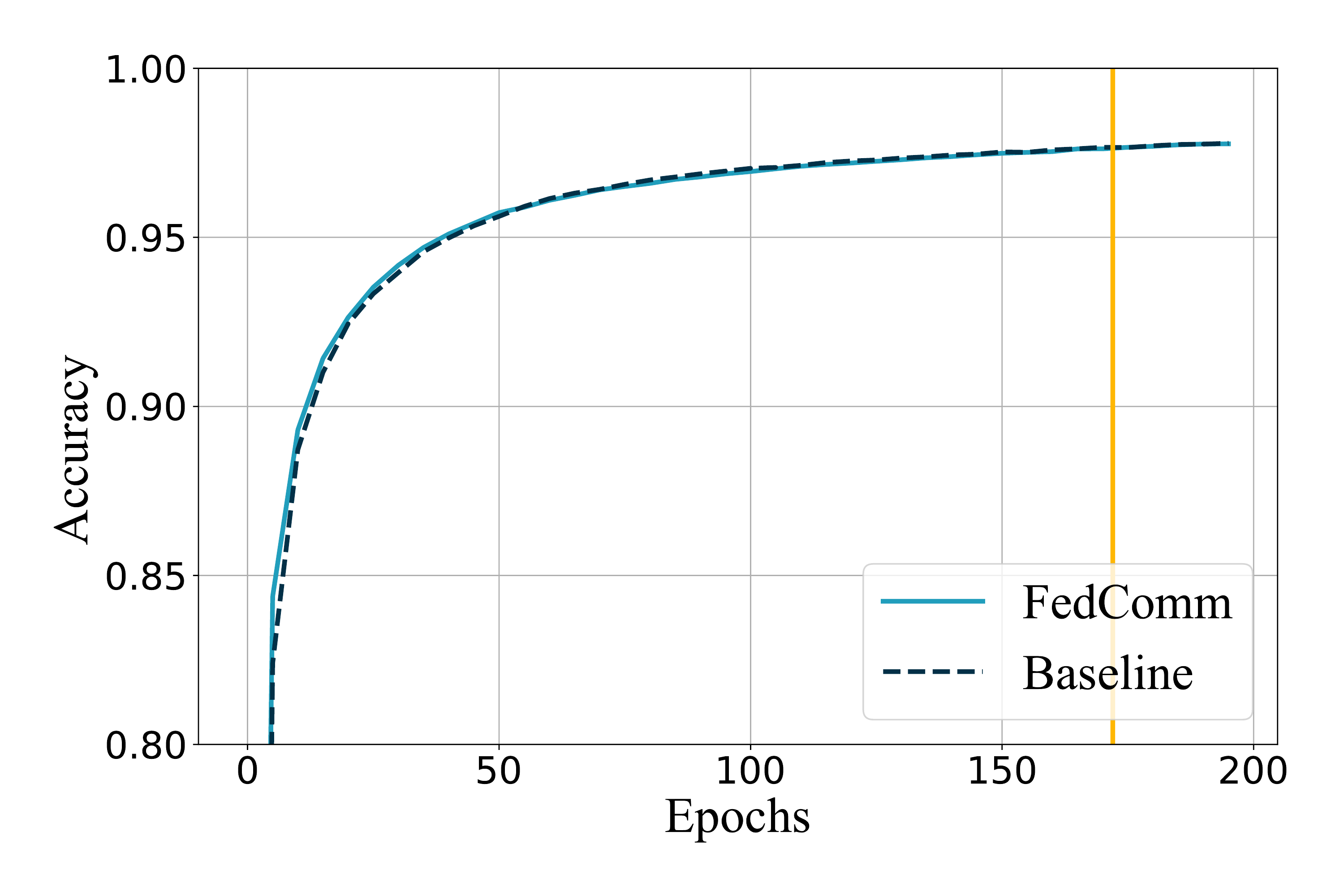}
             \caption{$\beta=0.99$, $\gamma = 0.1\sigma/\sqrt{P}$ \\ 10 senders, short message, MNIST\\ 100\% aggregated per round}
             \label{fig:10_full_stealthy_mnist}
        \end{subfigure}
        \hfill
        \begin{subfigure}{.3\textwidth}
	        \centering
            \includegraphics[width=\columnwidth]{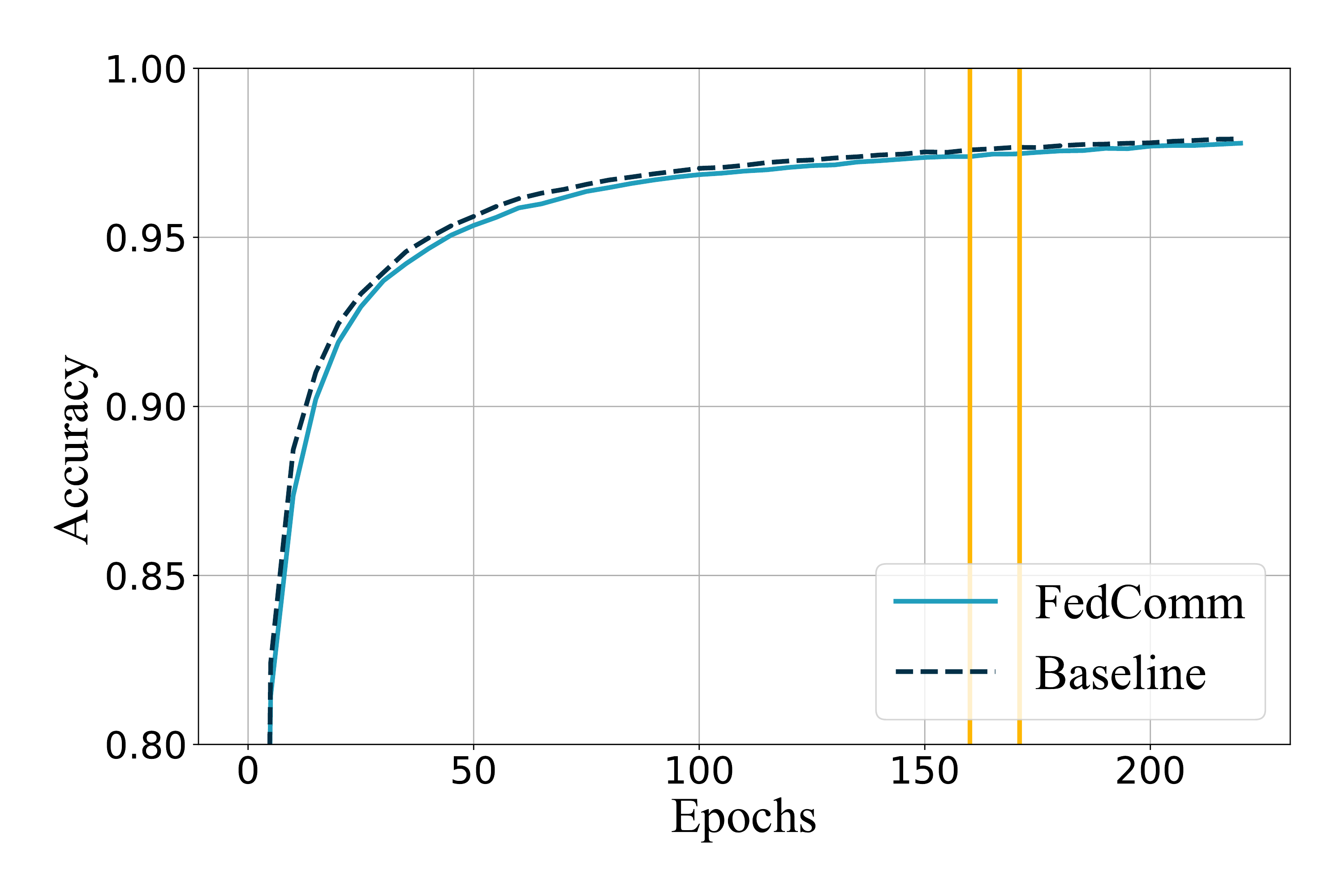}
            \caption{$\beta$=0, $\gamma = \sigma/\sqrt{P}$\\2 senders, 2 short messages, MNIST\\ 100\% aggregated per round.}
            \label{fig:non_stalthy_two_messages}
        \end{subfigure}
        \bigskip

    \centering
	    \begin{subfigure}{.3\textwidth}
            \centering
            \includegraphics[width=\columnwidth]{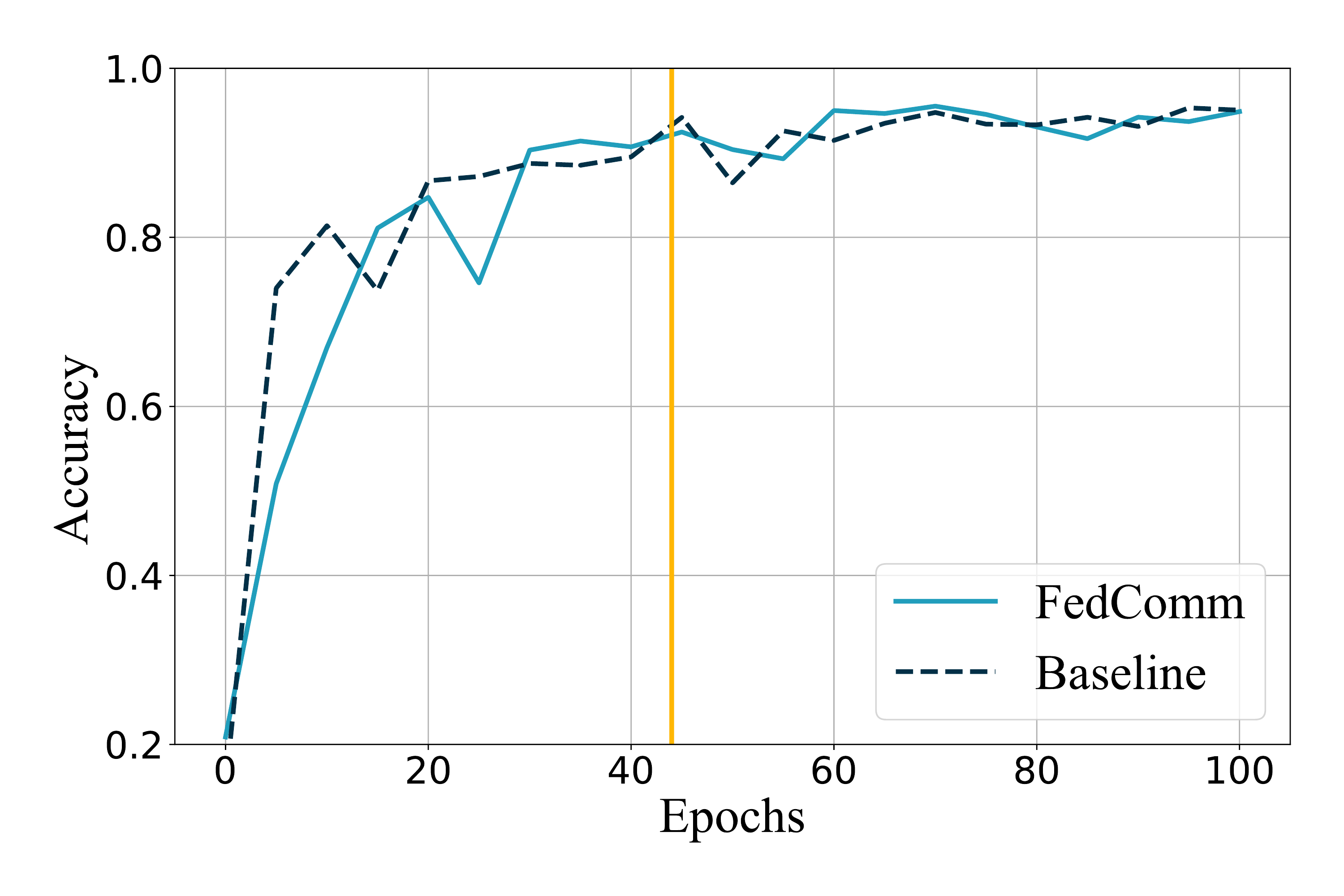}
             \caption{$\beta=1/\sqrt{2}$, $\gamma = \sigma/\sqrt{2P}$\\ 5 senders, short message, MNIST\\ 10\% aggregated per round}
             \label{fig:5_half_stealthy_mnist_10}
        \end{subfigure}
        \hfill
	    \begin{subfigure}{.3\textwidth}
	        \centering
            \includegraphics[width=\columnwidth]{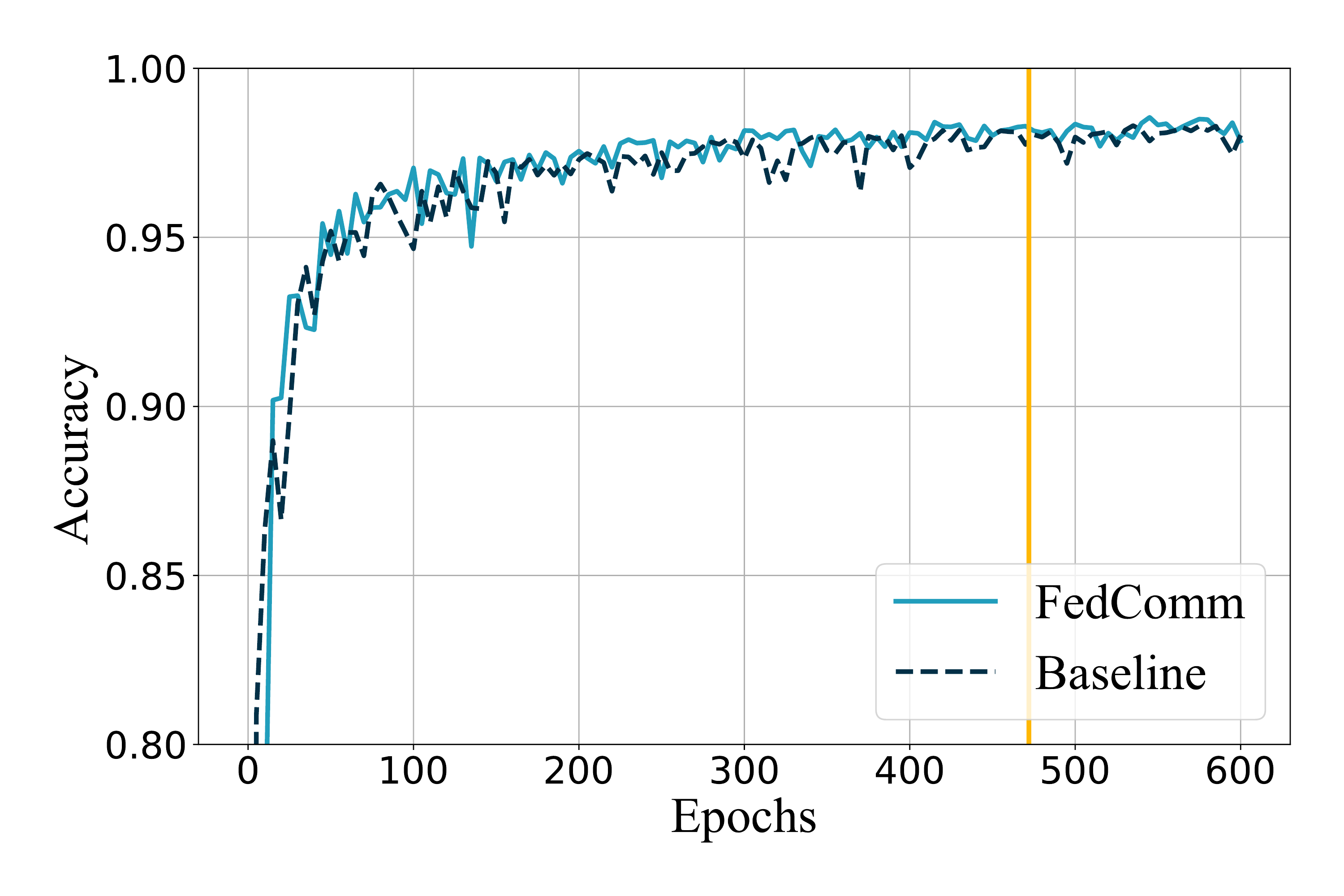}
            \caption{$\beta=0.99$, $\gamma = 0.1\sigma/\sqrt{P}$\\ 10 senders, short message, MNIST\\ 20\% aggregated per round.} 
            \label{fig:10_full_stealthy_20_percent}
        \end{subfigure}
        \hfill
	    \begin{subfigure}{.3\textwidth}
	        \centering
             \includegraphics[width=\columnwidth]{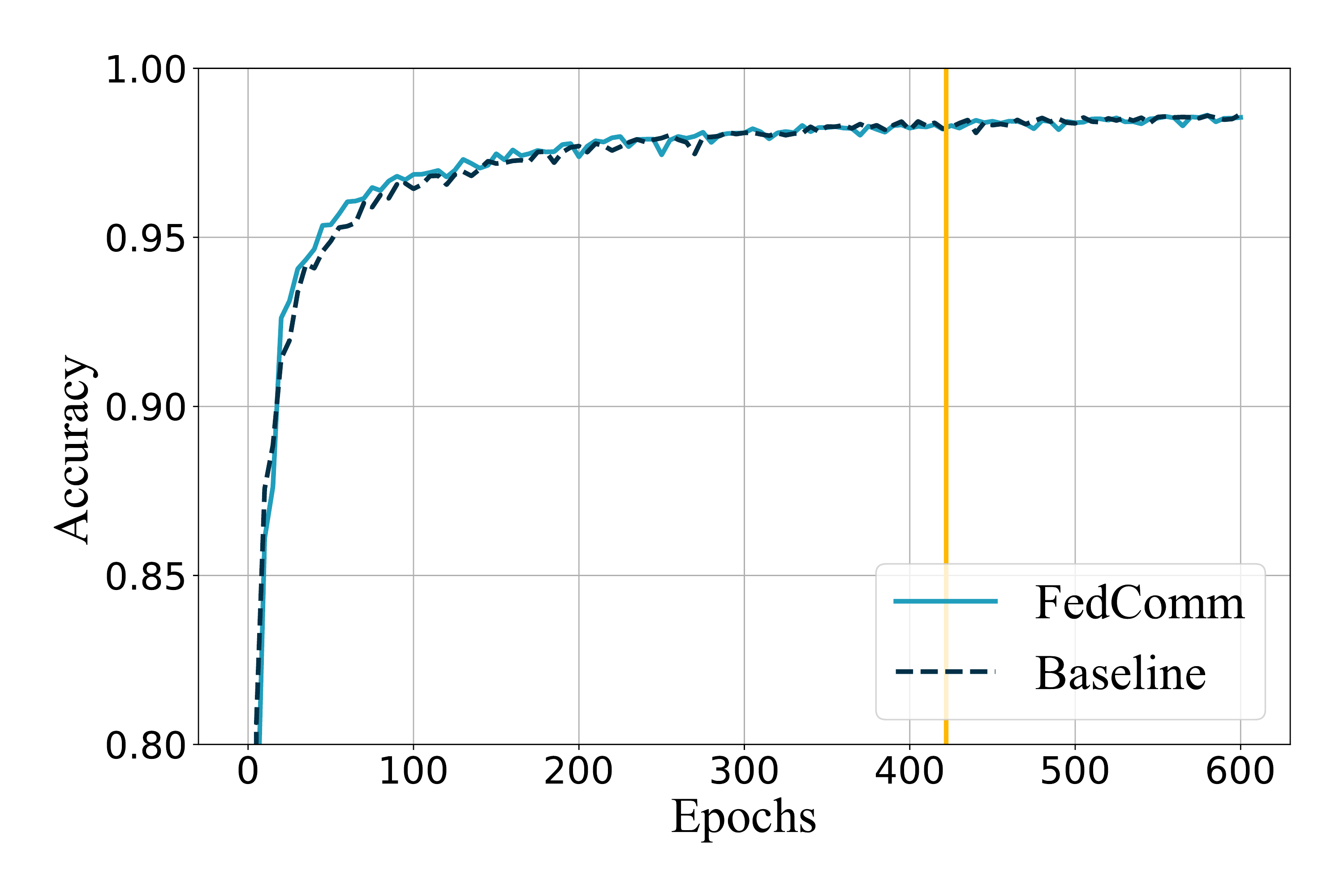}
            \caption{$\beta$=0, $\gamma = \sigma/\sqrt{P}$\\ 1 sender, short message, MNIST\\ 50\% aggregated per round} 
            \label{fig:non_stealthy_50_percent_aggregated}
        \end{subfigure}
        \bigskip

    \centering
	    \begin{subfigure}{.3\textwidth}
            \centering
            \includegraphics[width=\columnwidth]{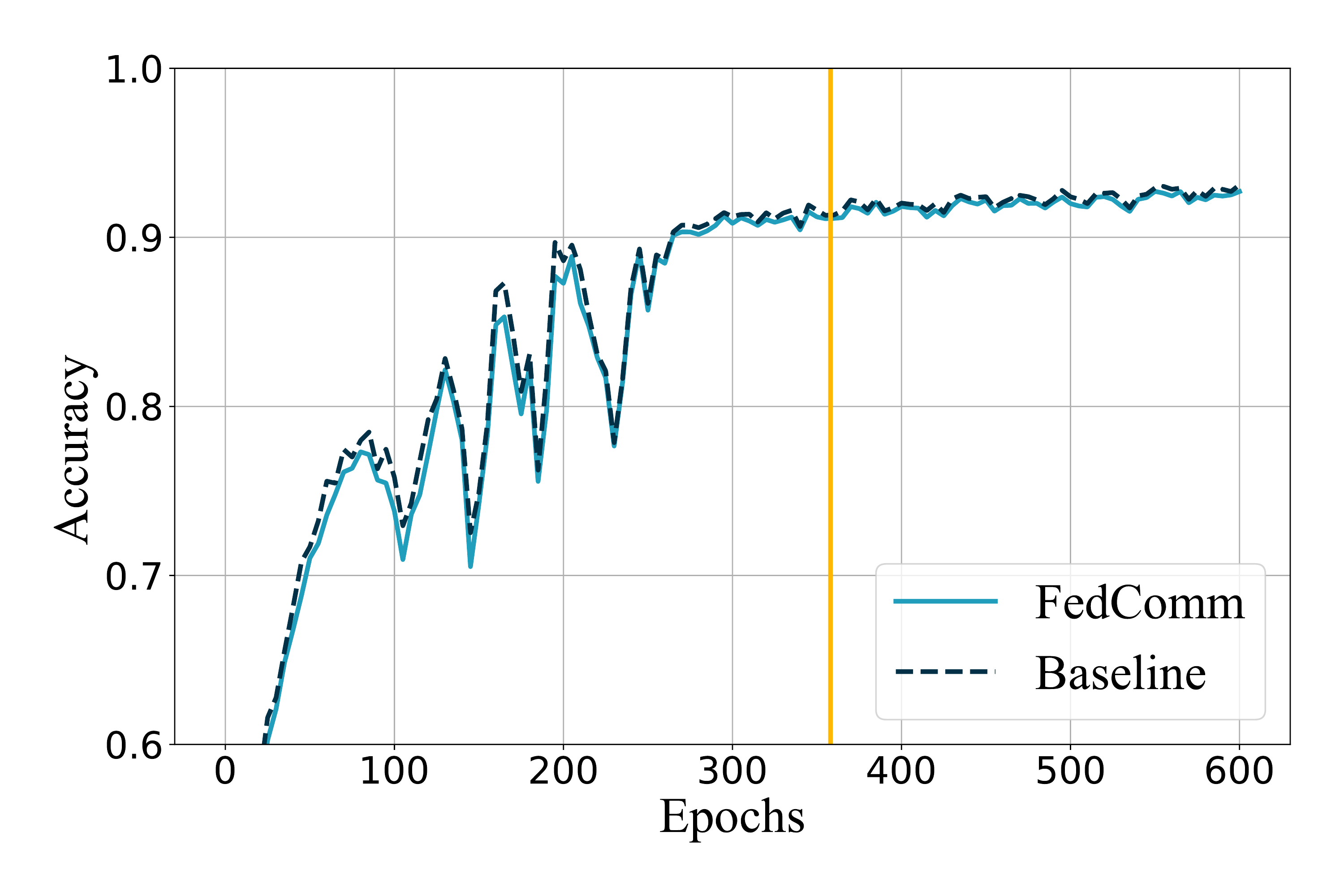}
            \caption{ $\beta=1/\sqrt{2}$, $\gamma = \sigma/\sqrt{2P}$\\5 senders, Long message, CIFAR10\\ 100\% aggregated per round} 
             \label{fig:half_stalthy_png}
        \end{subfigure}
        \hfill
	     \begin{subfigure}{.3\textwidth}
            \centering
            \includegraphics[width=\columnwidth]{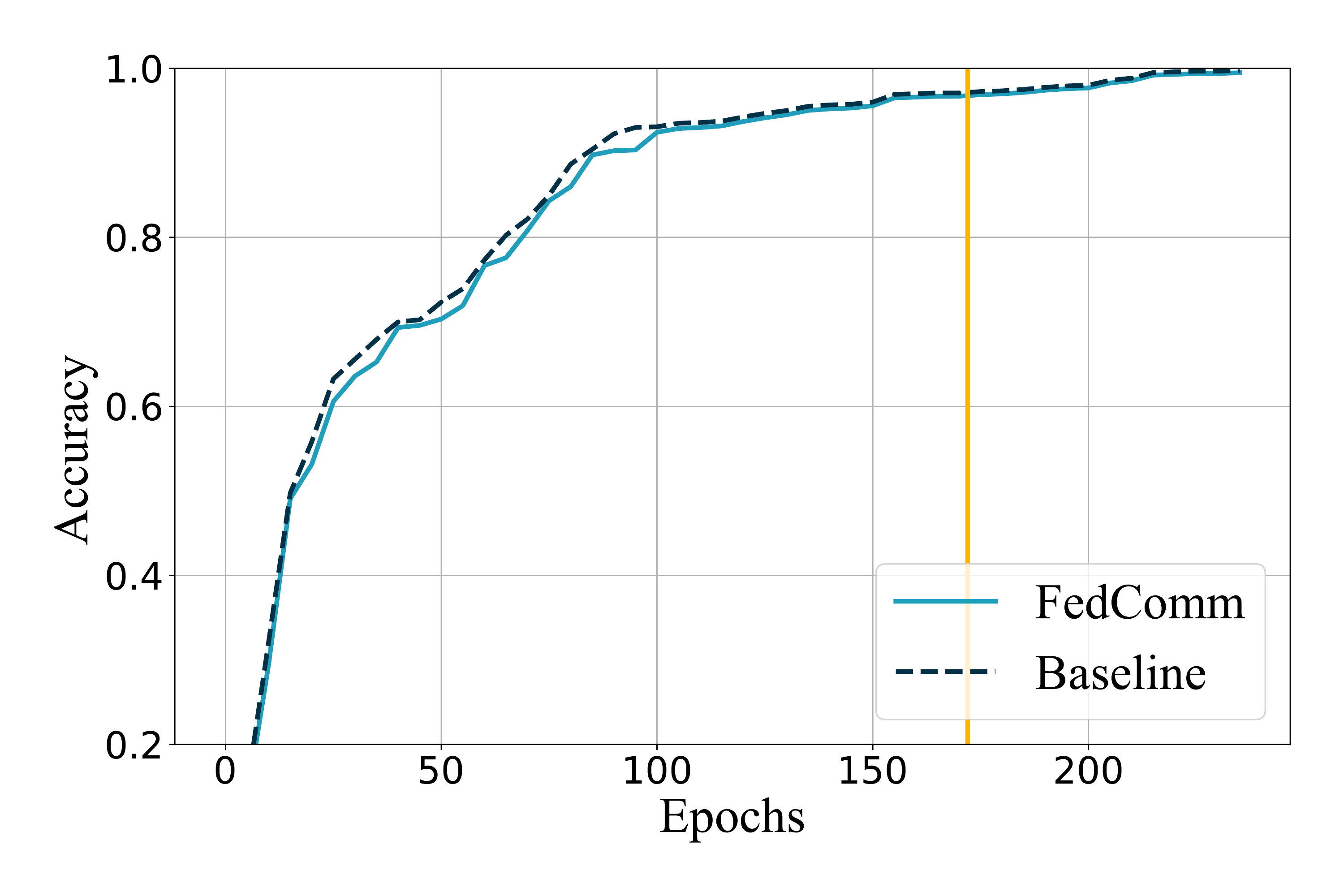}
            \caption{$\beta=0.99$, $\gamma = 0.1\sigma/\sqrt{P}$ \\10 senders, short message, ESC-50\\ 100\% aggregated per round.} 
             \label{fig:full_stealthy_audio}
        \end{subfigure}
        \hfill
	    \begin{subfigure}{.3\textwidth}
	        \centering
            \includegraphics[width=\columnwidth]{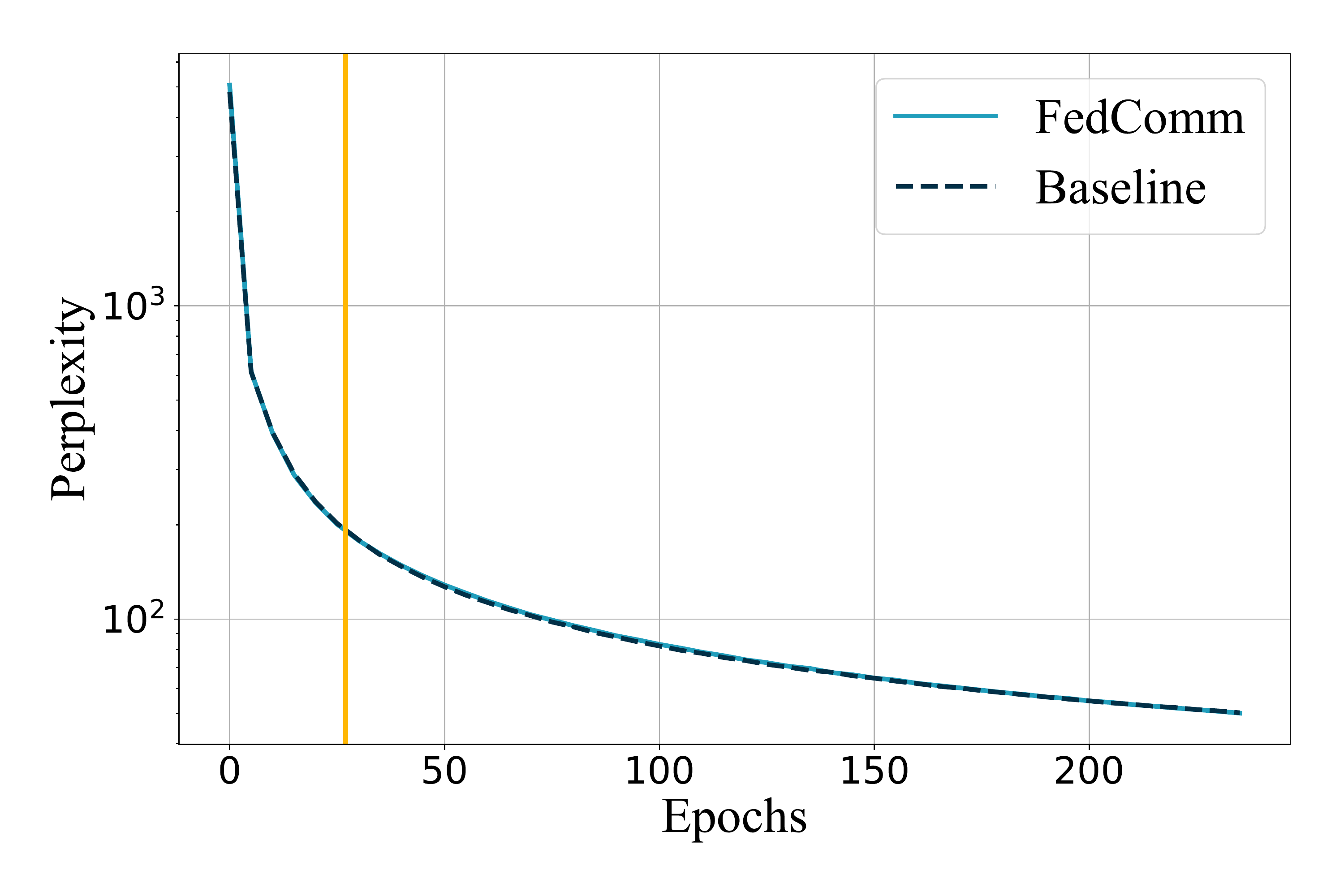}
            \caption{$\beta=0.99$, $\gamma = 0.1\sigma/\sqrt{P}$ \\10 senders, short message, WikiText-2\\ 100\% aggregated per round.} 
            \label{fig:full_stealthy_language}
        \end{subfigure}
        \bigskip

    \centering
	    \begin{subfigure}{.3\textwidth}
	        \centering
            \includegraphics[width=\columnwidth]{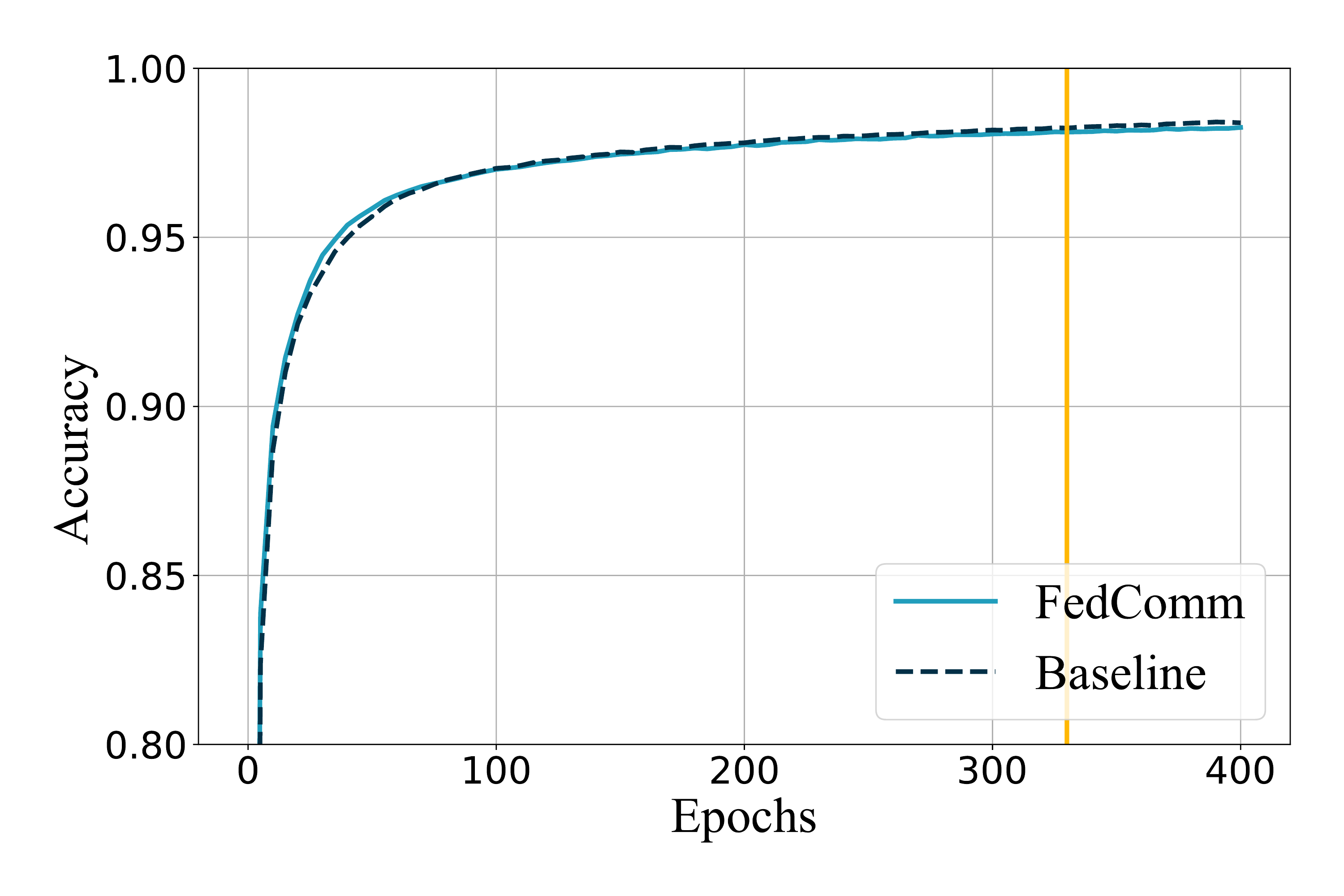}
            \caption{$\beta=1/\sqrt{2}$, $\gamma = \sigma/\sqrt{2P}$ \\1 sender, short message, MNIST\\ 100\% aggregated per round}
            \label{fig:1_half_stealthy_mnist}
        \end{subfigure}
        \hfill
        \begin{subfigure}{.3\textwidth}
	        \centering
            \includegraphics[width=\columnwidth]{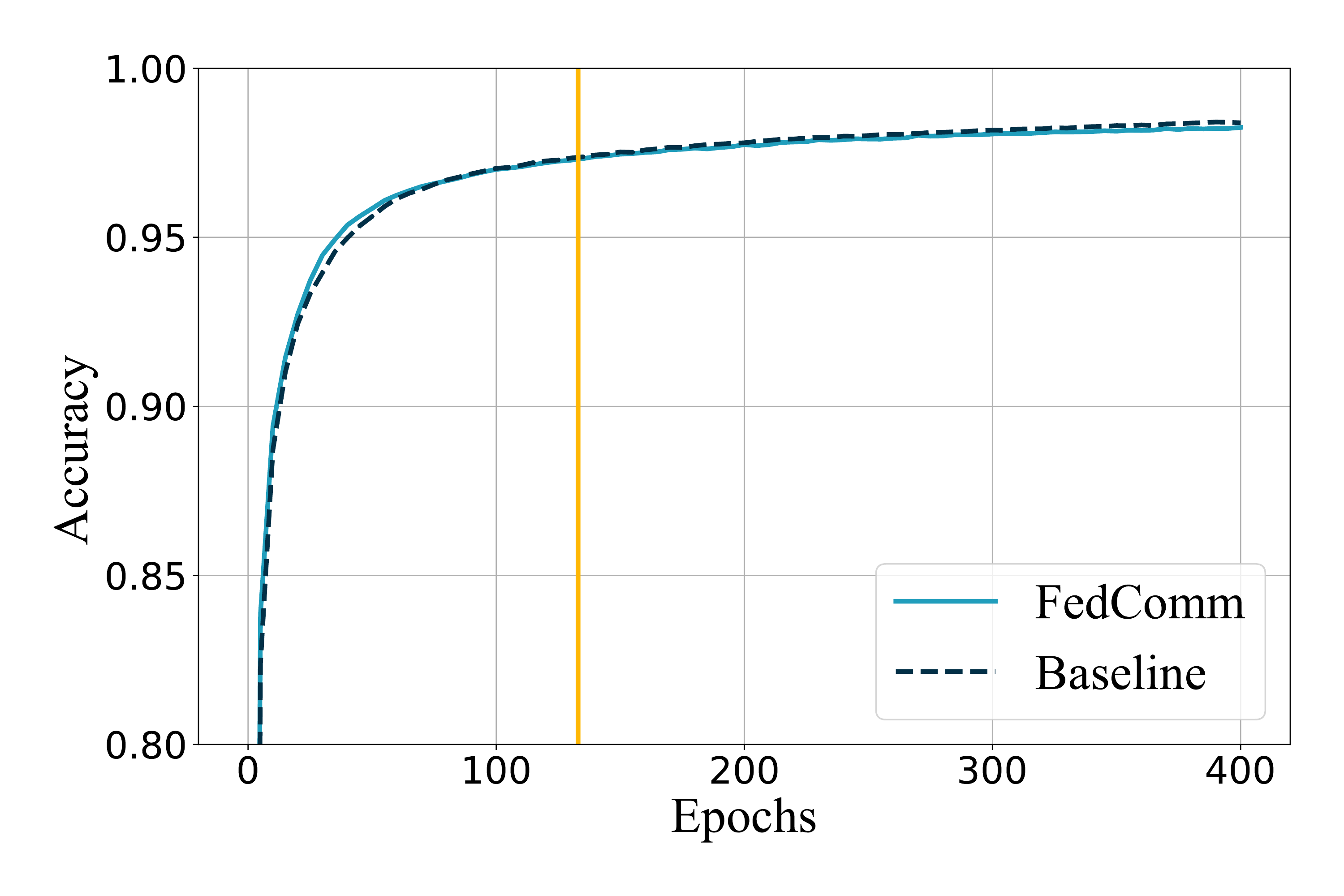}
            \caption{$\beta=1/\sqrt{2}$, $\gamma = \sigma/\sqrt{2P}$ \\2 senders, short message, MNIST\\ 100\% aggregated per round}
            \label{fig:2_half_stealthy_mnist}
        \end{subfigure}
        \hfill
        \begin{subfigure}{.3\textwidth}
	        \centering
            \includegraphics[width=\columnwidth]{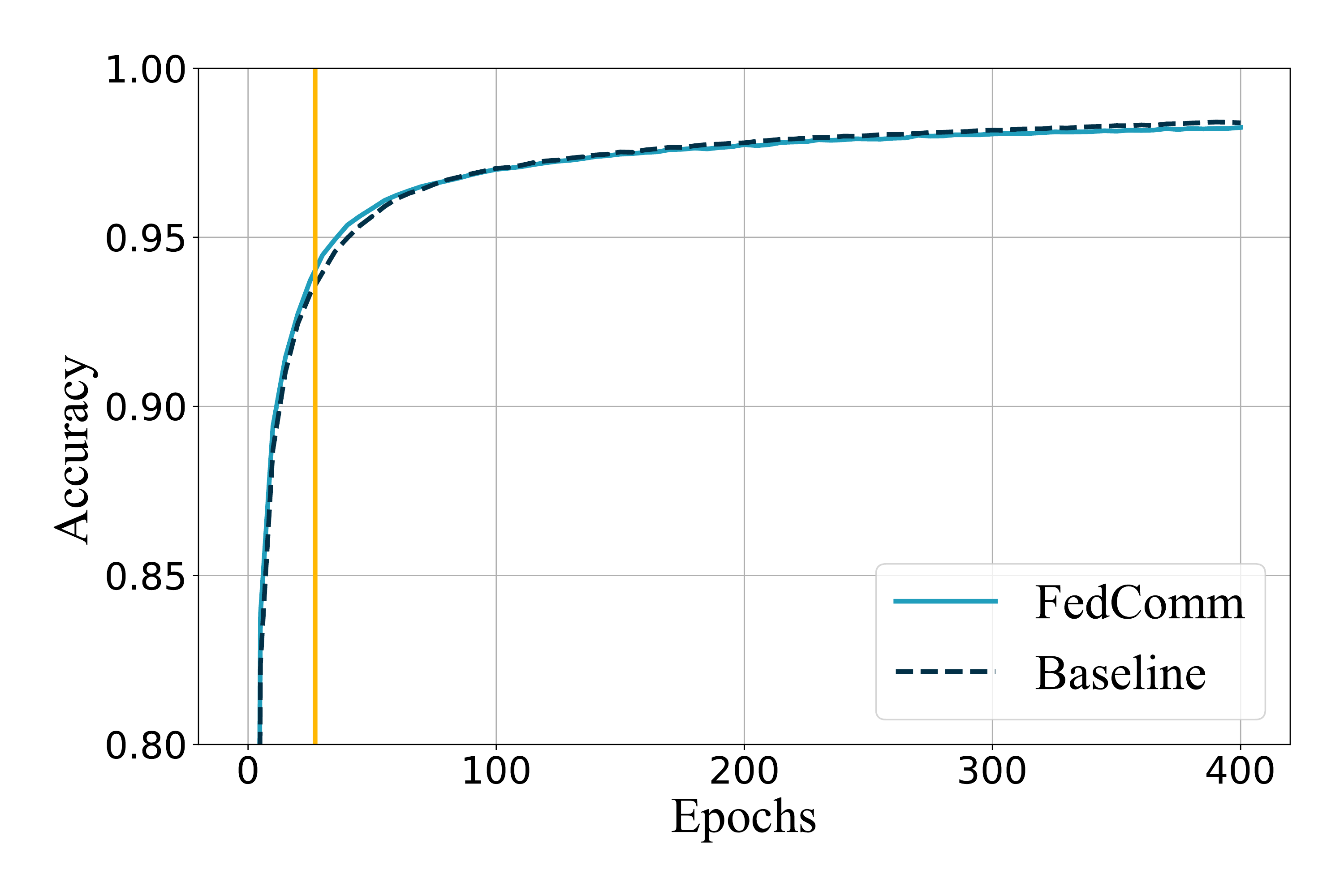}
            \caption{$\beta=1/\sqrt{2}$, $\gamma = \sigma/\sqrt{2P}$ \\4 senders, short message, MNIST\\ 100\% aggregated per round}
            \label{fig:4_half_stealthy_mnist}
        \end{subfigure}
        \bigskip
    \caption{The \name approach run on different FL setups. In each figure, the stealthiness parameters $\beta$ and \textbf{$\gamma$} are displayed with the number of senders, size of the message, dataset type, and the percentage of users selected at random to participate in each FL round. The vertical yellow line on each plot indicates the round where the message was able to be correctly received by the \textit{receiver}.} 
    \label{fig:baseline_vs_fedcomm_performance}
\end{figure*}

\subsection{Stealthiness}\label{sec:stealthiness}
During an FL epoch where the participants perform a round of training over their respective \emph{local} datasets, the gradients of the updated weights are uploaded to the parameter server, which, in turn, updates the global model. Because FL usually relies on secure aggregation~\cite{secureAggregation}, the parameter server is oblivious of the individual updates.
While typically, the parameter server of an FL scheme does not possess any tracking mechanism, in the following, we assume a \emph{hypothetical} scenario in which the parameter server can actually observe each of the uploaded updates. Additionally, the server is equipped with additional tooling for performing statistical analysis of the provided updates to detect and mitigate eventual anomalies. 
We position ourselves in such a scenario, as in this setting, we can adequately evaluate the stealthiness of \name and demonstrate that the transmitted messages remain undetected.

To demonstrate the stealthiness of \name when using different stealthiness parameters, we analyzed the distribution of the gradient updates that come from employing \name to transmit a \emph{message} versus the gradient updates when \name is not used. We compared the parameters' distribution of the gradient updates using the two-sample Kolmogorov-Smirnov (KS) statistical test. The two-sample KS test is used to test whether two samples come from the same distribution.
We performed this experiment on each architecture/payload combination (see section\ref{sec:exp_setup}).
In our experiments, according to the KS test, the distribution of gradient updates in the two cases, i.e., the regular gradient update and \name in non-stealthy mode, do not come from the same distribution.
On the other hand, the distribution of the gradient updates when using \name on \emph{full-stealthy mode} come from the same distribution as the regular gradient update according to the KS-test. Given that, we conclude that the full-stealthy \name transmission cannot be detected.
The impossibility of distinguishing between the distributions of a typical gradient update and the gradient update that contains the \emph{message} using \name in full-stealthy mode aligns with the theoretical results reported in Section~\ref{sec:fedcomm}, which show that using the stealthiness parameters $\beta=0.99$, $\gamma = 0.1\sigma/\sqrt{P}$ allows us to bury the \emph{message} in the gradients, thus making it undetectable. In this case, the information would be impossible to detect because the signal of the message would be 10dB under the power of the gradient.  Prior work~\cite{Torrieri2005Detection} in digital communication theory have shown that below 10 dB signal to noise ratio between the signal level (message) and noise levels (gradients) is not enough to detect (and separate) the wanted signal (message) from the noise (gradients).

For completeness, we visually depict this statistical comparison in Figure~\ref{fig:distribution_comparison}, in which we compare the distribution of the \name gradient updates in the two extreme cases; non-stealthy ($\beta$=0, $\gamma = \sigma/\sqrt{P}$) and full-stealthy ($\beta=0.99$, $\gamma = 0.1\sigma/\sqrt{P}$), with the regular gradient updates. 

In Figure~\ref{fig:non_stealthy_histogram}, the distribution of typical gradient updates after the local iterations (the light color) differs from the distribution of the updates where the \emph{message} is being transmitted in a non-stealthy manner.
When the \emph{message} is transmitted in non-stealthy mode, the parameter server can find out that something abnormal is happening and might even choose to discard that particular gradient update.

Figure~\ref{fig:full_stealthy_histogram} displays the distribution of the gradient updates after a typical local update, and the distribution of the same gradient updates with the \emph{message} transmitted using \name in full-stealthy mode. The two distributions are indistinguishable and in the eyes of the parameter server, nothing abnormal is happening, even when he is employing KS statistical test to determine whether the distributions are different or not.
To provide additional evidence of \name's ability to transmit a \emph{message} covertly, we compared the vector norm among all the participants (i.e., \emph{senders} and regular participants) gradient updates. Similar to the above results, the parameter server tries to detect anything unusual in the parameter updates from a particular participant compared to the parameter updates sent by the rest of the participants by employing a different measure, the norm of the gradient update. In this experiment, we used the Frobenius norm~\cite{norm_computation}. Figure~\ref{fig:norm_injection_round} shows that the norm of the gradient update of the \emph{sender} (highlighted in dark blue) is similar to the norm of the gradients updates of other participants.

\subsection{\name's impact on FL model performance}\label{sec:channel_impact}
To measure the impact on the performance of the resulting FL model when using the \name covert communication scheme, we ran different experiments on a variety of tasks (MNIST, CIFAR10, ESC-50, WikiText-2) and a variety of DNN architectures (see Section~\ref{sec:exp_setup}). In this way, we also empirically evaluate the generality of \name (i.e., domain- and DNN-architecture independent).
We fixed the number of participants in the FL scheme to 100 and considered the following cases in terms of the percentage of participants randomly selected to update the parameters in each round (10\%, 20\%, 50\%, 100\%). To show that \name does not impact the performance of the FL scheme, we performed baseline runs with the same setup as the one in which we used \name to transmit the message, and we display those results in Figure~\ref{fig:baseline_vs_fedcomm_performance}. Each plot presents the FL baseline training accuracy against the training accuracy when \name is employed. The message is transmitted on each round the \emph{sender} is selected for participating (i.e. when we use 100\% this means the \emph{sender} is always selected and transmits the message in each round). The vertical yellow line in each plot shows the FL global round $T$ on which the \emph{message} is correctly received by the \emph{receiver}.
To display the effect on model performance when the \emph{sender} employs different levels of \name stealthiness,
Figure~\ref{fig:one_non_stealthy_mnist} and Figure~\ref{fig:10_full_stealthy_mnist} display the model performances when using \name non-stealthy (\ref{fig:one_non_stealthy_mnist}) and \name full-stealthy~(\ref{fig:10_full_stealthy_mnist}) level to transmit the \emph{message}. On both cases we can see that the performance of the learned model when \name is used is similar to the performance of the learned model when \name is not used.
Another essential benefit that results from the use of spread-spectrum channel coding techniques
is that multiple \emph{senders} can concurrently send their respective \emph{messages} to their respective target \emph{receivers}. Figure~\ref{fig:non_stalthy_two_messages} showcases an experiment in which two \emph{senders} transmit two different text messages (i.e., \emph{hello world!} and \emph{The answer is 42!}) to their respective \emph{receivers}. The performance of the learned model is unaffected on each FL round, and both \emph{messages} are correctly delivered to their respective \emph{receivers}.
As previously mentioned in Section~\ref{sec:fedcomm}, we tested the case in which a subset of the total participants were selected at random to update the global model in each FL round. We assessed the impact of this approach on \name's ability to transmit the \emph{message} in Figures~\ref{fig:5_half_stealthy_mnist_10},~\ref{fig:10_full_stealthy_20_percent},~\ref{fig:non_stealthy_50_percent_aggregated}. These figures show that the baseline and \name's training accuracy is still closely comparable even when a limited number of participants is selected for averaging (i.e., 10\%, 20\% and 50\%). 

Figures~\ref{fig:half_stalthy_png},~\ref{fig:full_stealthy_audio},~\ref{fig:full_stealthy_language} evaluate \name's performance when it is applied on different tasks and different DNN architectures. Figure~\ref{fig:half_stalthy_png} shows that \name can transmit the long message in under 400 global rounds while simultaneously training the VGG~\cite{Simonyan14verydeep} network on the CIFAR10 image-recognition task. Figure~\ref{fig:10_full_stealthy_20_percent}, displays the baseline vs. \name's performance on the ESC-50~\cite{piczak2015dataset} audio classification task. The performance of the learned model on each round is not affected by the ongoing covert communication powered by \name. Figure~\ref{fig:full_stealthy_language} displays the baseline vs. \name performance on a language-modeling task, WikiText-2, using a LSTM-based recurrent NN. Different from other plots, the performance assessment on this task compared the perplexity of the learned models on each FL round. Perplexity measures how well a probability model predicts a sample, and in our case, the language model is trying to learn a probability distribution over the WikiText-2~\cite{wikitext_dataset} dataset. Even in this case, the performance of the FL scheme is not impacted at all by having \name transmitting a \emph{message} alongside the learning process.

Figures~\ref{fig:1_half_stealthy_mnist},~\ref{fig:2_half_stealthy_mnist},~\ref{fig:4_half_stealthy_mnist} show \name's results when using the same settings (task, message and stealthiness level) with changes to the number of simultaneous \emph{senders} of the same message and demonstrate how this scenario impacts the \emph{message} delivery time. We elaborate on this in the next section where we discuss about the message delivery time of \name.

\subsection{Message delivery time}\label{sec:message_delivery_time}

Having shown that employing \name we can covertly deliver a \emph{message} without impacting the ability of the FL scheme to learn a high performing ML model, we focused on measuring the time (i.e. the latency in terms of FL rounds) it takes a \emph{message} to be delivered to its intended \emph{receiver}.
From the various FL configuration experiments demonstrated in Figure~\ref{fig:baseline_vs_fedcomm_performance}, message delivery time varies according to the number of \textit{senders} in the network and their \textit{stealthiness} levels.

Typically, an FL scheme either runs in a continuous learning fashion (i.e., the learning never stops) or stops when the gradient updates can no longer improve the model. 
To display the potential of \name, we assume the latter case, which is also our worst-case scenario because it requires \name to transmit and deliver the \emph{message} to the \emph{receiver} before the FL procedure converges (i.e., the training stops). In Figure~\ref{fig:baseline_vs_fedcomm_performance}, the vertical line indicates the global round $T$ in which the \textit{receiver} correctly received the message. On every FL run, the model performance continues improving, even after the FL round when the message is received, so the FL execution does not stop before the message is received.

\begin{figure}[htp]
    \centering
    \includegraphics[width=.8\columnwidth]{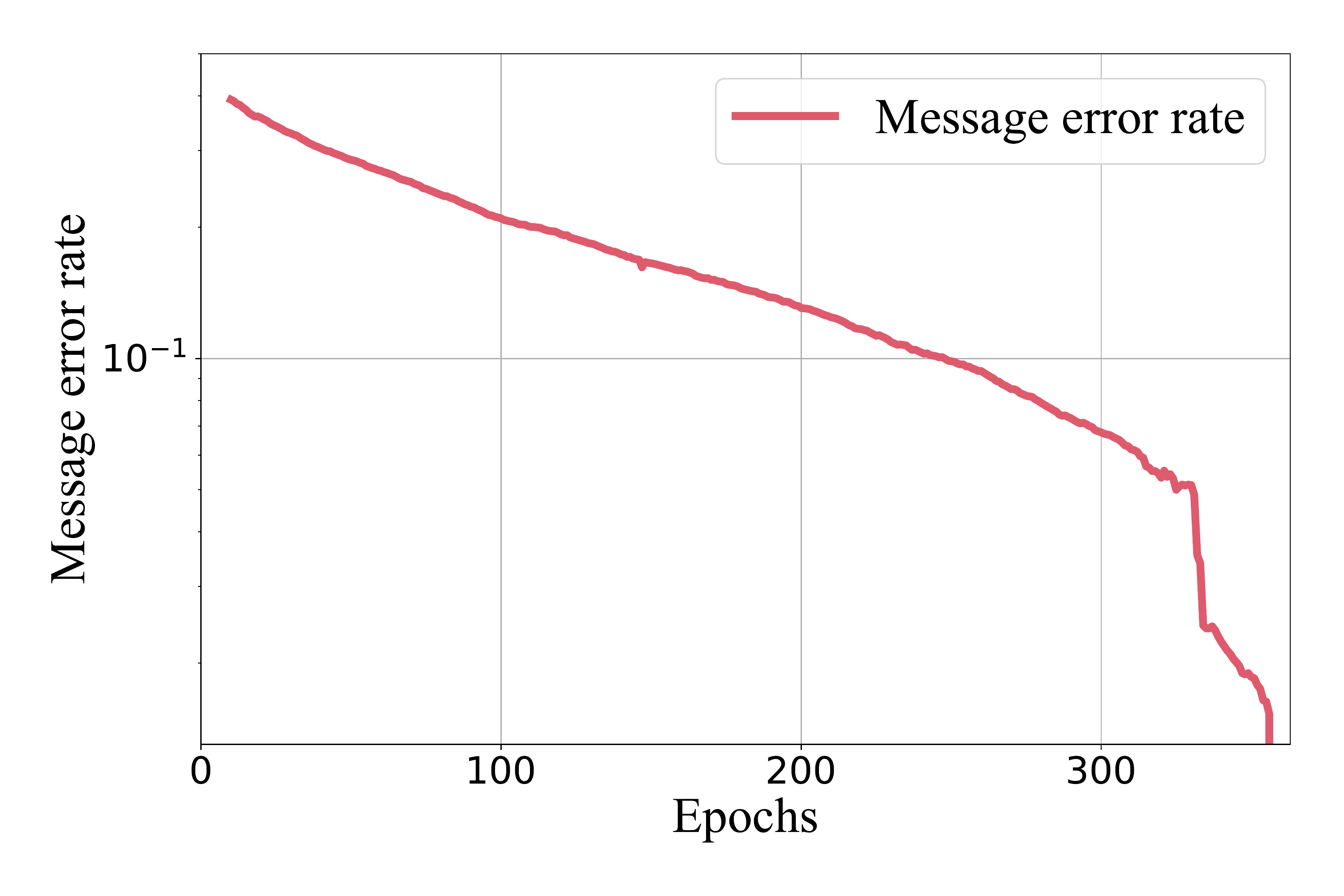}
    \caption{Error rate until the \emph{message} can be retrieved from the receiver. $\beta=1/\sqrt{2}$, $\gamma = \sigma/\sqrt{2P}$, 5 senders, Long message, CIFAR10, 100\% aggregated per round.} 
    \label{fig:error_rate}
\end{figure}

The message delivery time drops significantly as the number $M$ of \textit{senders} who send the message concurrently increases. In Section~\ref{sec:fedcomm}, we showed that the number of iterations drops with $M^2$. We highlight this observation on the experiments displayed on Figures~\ref{fig:1_half_stealthy_mnist},~\ref{fig:2_half_stealthy_mnist},~\ref{fig:4_half_stealthy_mnist}, which show the message delivery time when using 1, 2, and 4 \textit{senders} with the stealthiness parameter $\beta=1/\sqrt{2}$, $\gamma = \sigma/\sqrt{2P}$, and on each round, 100\% of participants are selected by the \textit{parameter server} for averaging. 
The vertical line in Figure~\ref{fig:1_half_stealthy_mnist} shows that fewer than 400 global rounds are needed for the \textit{receiver} to be able to correctly decode the message when one \textit{sender} is used. 
According to the calculations, two \textit{senders} would require roughly $400/2^{2}=100$ and four \textit{senders} would require roughly $400/4^{2}=25$ FL rounds for the message to be correctly decoded.
Figure~\ref{fig:2_half_stealthy_mnist} and Figure~\ref{fig:4_half_stealthy_mnist} show that with 2 and 4 concurrent \textit{senders}, the \textit{receiver} can decode the message in under 120 and 30 rounds respectively. 
The small mismatch between the theory (the number of iterations is reduced by $M^2$) and practice (slightly slower rate of decrease, especially from 1 to 2 \emph{senders} in Figures~\ref{fig:1_half_stealthy_mnist} and~\ref{fig:2_half_stealthy_mnist}) can be due to several factors. First, the variance of the gradients reduces as the number of iterations increases, so there is more noise present in the first few iterations. Second, the gradients for each \emph{sender} are different, so all of them might not be adding the same amount of information in each iteration. Third, the natural stochasticity in FL training, as each user has a different gradient in each iteration that depends on all the other gradients in previous iterations. We can also see in Figures~\ref{fig:one_non_stealthy_mnist} and~\ref{fig:10_full_stealthy_mnist} that ten stealthy users take a time that is of the same order of magnitude as one non-stealthy user, as $\delta^2$ and $M^2$ would cancel each other out in the predicted number of iterations, which is also predicted by the theory in Section~\ref{sec:fedcomm}. 

Finally, to highlight the progress by which the \emph{message} becomes visible in the global model, Figure~\ref{fig:error_rate} shows the decrease in \emph{message} error rate after each FL round.
This error rate is calculated as the portion of the \emph{message} that is received by the \emph{receiver} after each FL round. Note that the use of LPDC codes causes the error rate to drop rapidly to zero in the last few iterations of \name.
\leavevmode \\
\textbf{Validating the Gaussian assumption.}
In Section~\ref{sec:fedcomm}, we assumed that the gradient updates in every round are equally distributed, as a zero-mean Gaussian with a constant variance $\sigma^2$ for tractability of the theoretical analysis. In practice, the gradients are not Gaussian distributed and do not have the same standard deviation for every user at each iteration. Hence the resulting noise on our covert communication channel would not be Gaussian distributed and this would have an impact on the number of iterations that are needed to be able to decode the message. 
In Figure~\ref{fig:one_non_stealthy_mnist}, we can see that we need 120 iterations before the message can be retrieved. If we apply the theory developed in Section~\ref{sec:fedcomm}, we should expect to decode the message after two iterations. We have done two experiments to understand where this deviation comes from. First, we recorded the power of all the gradient updates in each iteration and for each user. Then we repeated the FL procedure, substituting the gradient updates for each user with Gaussian noise but using the different variance values that were previously collected\footnote{In this experiment, there is no learning happening. We just want to see when the message is correctly retrieved.}. In this case, we were able to detect the message within 30 rounds.
Second, if we use the same Gaussian distribution for all the users and for all the iterations, the resulting noise is Gaussian distributed, and we recover the message in two iterations, as predicted by the theory. 

Both the LDPC and the CDMA detector rely on the resulting noise being Gaussian for optimal performance. The degradation observed in our experiments can be mitigated if we had the exact distribution of the noise that our CDMA communication is suffering. But this distribution would depend on the architecture of the DNN, the data that each user has, and each iteration of the FL procedure. It would be impossible to theoretically predict the number of needed iterations. In this work, we have focused on showing that the message can be retrieved and have left as further work designing the optimal detector. 

%% file: sections/08-discussion.tex
\section{Potential Countermeasures}
\label{sec:discussion}
In this section we analyze possible approaches that can be employed as countermeasures to \name,
and discuss the extent to which these countermeasures can impact the performance of the covert communication channel.

\leavevmode \\
\textbf{Parameter Pruning}
 is a technique that is commonly used to reduce the size of a neural network while attempting to retain a similar performance as the non-pruned counterpart. Parameter pruning comprises of removing unused or least-used neurons of a neural network. Detecting and pruning these neurons requires a dataset that represents the whole population on which this NN was trained. The least activated neurons are identified and pruned by iteratively querying the model. 
Pruning is not applicable in federated learning because neither the parameter server nor the participants possess a dataset representing the whole population. Going against the FL paradigm, we assume that the parameter server has such a dataset. Suppose the parameter server performs the pruning during the learning phase. In that case, the \textit{sender} will find out when it downloads the next model update and can re-transmit the message using as target weights the parameters of this new architecture. 
If a \textit{sender} wants to increase the chances that a message will not be disrupted by pruning, he can analyze the model updates to discover the most used parameters, which are less likely to be pruned, and use that subset of parameters to transmit the message.
\leavevmode \\
\textbf{Gradient Clipping.} is a technique to mitigate the exploding gradient problem in DNNs~\cite{zhang2019gradient}. Typically, gradient clipping introduces a pre-determined threshold and then scales down the gradient norms that exceed the threshold to match the norm, introducing a bias in the resulting values of the gradient, which helps stabilize the training process. In a federated learning scenario, the aggregator could employ gradient clipping on participants’ gradients.
Under this setting, we performed gradient clipping at the aggregator using norm ranges of $[0.5, 0.6, \dots, 1.0]$. On each global round, the aggregator scales each of the received gradients by the participants to match the pre-set norm. From our experimental evaluation, we observe that this method incurs no penalty on \name's ability to transmit the message.

%% file: sections/09-related_work.tex
\section{Related Work}\label{sec:related_work}

Recent years have seen an increasing and constantly evolving pool of attacks against deep-learning models, and also FL is shown to be susceptible to adversarial attacks~\cite{kairouz2019advances, jere2020taxonomy, lyu2020threats}.
For instance, while FL is designed with privacy in mind~\cite{mcmahan2017communication, shokri2015privacy}, attacks such as property-inference~\cite{ateniese2015hacking, melis2019exploiting, zhang2021leakage}, model inversion~\cite{fredrikson2015model}, and other generative adversarial network based reconstruction attacks~\cite{hitaj2017deep}, have shown that the privacy of the users participating in the FL protocol can be compromised too.
One of the first property-inference attacks is~\cite{ateniese2015hacking} from Ateniese~et al., which shows how an adversary with white-box access to an ML model can extract valuable information about the training data. Fredrikson~et al.~\cite{fredrikson2015model} extended the work in~\cite{ateniese2015hacking} by proposing a model-inversion attack that exploits the confidence values revealed by ML models.
Along this line of work, Song~et al.~\cite{song2017machine} demonstrated that it is possible to design algorithms that can embed information about the training data into the model (i.e., backdooring), and how it is possible to extract the embedded information from the model given only black-box access. 
Ganju~et al.~\cite{ganju2018property} extended~\cite{ateniese2015hacking} by crafting a property-inference attack against fully connected neural networks, exploiting the fact that fully-connected neural networks are invariant under permutation of nodes in each layer.
On the other hand, Zhang~et al.~\cite{zhang2021leakage} extended the above-mentioned property-inference attacks~\cite{ateniese2015hacking, ganju2018property} in the domain of multi-party learning by devising an attack that can extract the distribution of other parties' sensitive attributes in a black-box setting using a small number of inference queries.
Melis~et al.~\cite{melis2019exploiting} crafted various membership-inference attacks against FL protocol, under the assumption that the participants upload their weights to the parameter server after each mini-batch instead of after a local training epoch.
Hitaj~et al.~\cite{hitaj2017deep} demonstrated that a malicious participant in a collaborative deep-learning scenario can use generative adversarial networks (GANs) to reconstruct class representatives.
On the other hand, Bhagoji~et al.~\cite{bhagoji19icml} presented a model-poisoning attack that can poison the global model while ensuring convergence by assuming that the adversary controls a small number of participants of the learning-scheme.

\name is not an attack towards the federated learning protocol; we aim to transmit a hidden message within the model's updated parameters, without impeding the learning process~(Section~\ref{sec:channel_impact}).

\leavevmode \\
\textbf{Backdooring Federated Learning.}
Backdoors are a class of attacks~\cite{Gu2017BadNetsIV, Chen2017TargetedBA} against ML algorithms where the adversary manipulates model parameters or training data in order to change the classification label given by the model to specific inputs. 
Bagdasaryan~et al.~\cite{bagdasaryan2020backdoor} were the first to show that FL is vulnerable to this class of attacks. Simultaneously, Wang~et al.~\cite{wang2020attack} presented a theoretical setup for backdoor injection in FL, demonstrating how a model that is vulnerable to adversarial attacks is, under mild conditions, also vulnerable to backdooring.
Because this class of attacks is particularly disruptive, during the years, many mitigation techniques~\cite{Chen2017TargetedBA, Liu2018FinePruningDA, Tran2018SpectralSI, Chen2019DetectingBA} have been proposed. Burkhalt~et al.~\cite{burkhalter2021rofl} presented a systematic study to assess the robustness of FL, extending FL's secure aggregation technique proposed in~\cite{secureAggregation}. Burkhalt~et al.~\cite{burkhalter2021rofl} integrated a variety of properties and constraints on model updates using zero-knowledge proof, which is shown to improve FL's resilience against malicious participants who attempt to backdoor the learned model.
With a similar end goal to ours (i.e transmitting a message in the FL setting covertly), Costa~et al.~\cite{covertChannelAttacksFL} aim at exploiting backdooring of DNNs in FL to encode information that can be retrieved by observing the predictions given by the global model at that particular round.
 We highlight three-major advantages of \name when compared to~\cite{covertChannelAttacksFL}:
1) The covert channel of~\cite{covertChannelAttacksFL} relies on backdooring FL~\cite{bagdasaryan2020backdoor}, requiring specific tailoring to each domain. \name does not rely on backdooring FL, and more importantly \name is domain-independent. 
2) Bagdasaryan~et al.~\cite{bagdasaryan2020backdoor} emphasize that the model needs to be close to convergence to achieve successful backdoor injection. \name is not bound by such restrictions.
3) Work on backdoor detection in FL~\cite{Chen2019DetectingBA, Liu2018FinePruningDA, Tran2018SpectralSI}, can detect the covert channel introduced by~\cite{covertChannelAttacksFL}.
\name is not a backdooring attack and does not attempt to alter in any way the behaviour of the learned model. 
In \name's full-stealth mode (Section~\ref{sec:fedcomm}), the gradient updates of the \emph{sender} participant do not differ from the updates of other participants of the FL scheme. As such, backdooring defenses cannot prevent \name covert communication. 

%% file: sections/10-conclusions.tex
\section{Conclusions}\label{sec:conclusions}
This work proposes \name, an innovative covert communication technique that uses the FL scheme as a communication channel. We employ a combination of Code-Division Multiple Access spread-spectrum and LDPC error correction techniques to transmit the desired message correctly and covertly during the ongoing FL procedure. The sender hides the message in the gradient of the weight parameters of the local neural network architecture being trained and then uploads those parameters to the parameter server where they are aggregated.
\name does not introduce any particular artifact during the learning process, such as supplying inconsistent inputs, attempting to poison the model, or providing updates that differ significantly from other participants.
 We show that transmitting the message using \name cannot be detected by the global parameter server even if the server could observe individual gradient updates. 
Furthermore, \name does not hamper the FL scheme. We empirically show this by observing the accuracy and the loss of the global model at every round when the communication occurs. 

\section{Acknowledgments}
This work was partially supported by project SERICS (PE00000014) under the MUR National Recovery and Resilience Plan funded by the European Union - NextGenerationEU.

%% file: main.bbl
\begin{thebibliography}{10}
\providecommand{\url}[1]{#1}
\csname url@samestyle\endcsname
\providecommand{\newblock}{\relax}
\providecommand{\bibinfo}[2]{#2}
\providecommand{\BIBentrySTDinterwordspacing}{\spaceskip=0pt\relax}
\providecommand{\BIBentryALTinterwordstretchfactor}{4}
\providecommand{\BIBentryALTinterwordspacing}{\spaceskip=\fontdimen2\font plus
\BIBentryALTinterwordstretchfactor\fontdimen3\font minus
  \fontdimen4\font\relax}
\providecommand{\BIBforeignlanguage}[2]{{%
\expandafter\ifx\csname l@#1\endcsname\relax
\typeout{** WARNING: IEEEtranS.bst: No hyphenation pattern has been}%
\typeout{** loaded for the language `#1'. Using the pattern for}%
\typeout{** the default language instead.}%
\else
\language=\csname l@#1\endcsname
\fi
#2}}
\providecommand{\BIBdecl}{\relax}
\BIBdecl

\bibitem{Torrieri2005Detection}
\BIBentryALTinterwordspacing
\emph{Detection of Spread-Spectrum Signals}.\hskip 1em plus 0.5em minus
  0.4em\relax Boston, MA: Springer US, 2005, pp. 387--408. [Online]. Available:
  \url{https://doi.org/10.1007/0-387-22783-0_7}
\BIBentrySTDinterwordspacing

\bibitem{DziedzicBeyondFederation21}
D.~Adam, Choquette-Choo, C.~A., D.~Natalie, and N.~Papernot, ``Beyond
  federation: collaborating in ml with confidentiality and privacy,''
  \url{http://www.cleverhans.io/2021/05/01/capc.html}, 2021.

\bibitem{ateniese2015hacking}
G.~Ateniese, L.~V. Mancini, A.~Spognardi, A.~Villani, D.~Vitali, and G.~Felici,
  ``Hacking smart machines with smarter ones: How to extract meaningful data
  from machine learning classifiers,'' \emph{International Journal of Security
  and Networks}, vol.~10, no.~3, pp. 137--150, 2015.

\bibitem{bagdasaryan2020backdoor}
E.~Bagdasaryan, A.~Veit, Y.~Hua, D.~Estrin, and V.~Shmatikov, ``How to backdoor
  federated learning,'' in \emph{International Conference on Artificial
  Intelligence and Statistics}.\hskip 1em plus 0.5em minus 0.4em\relax PMLR,
  2020, pp. 2938--2948.

\bibitem{http_covert_channels}
M.~Bauer, ``New covert channels in http: Adding unwitting web browsers to
  anonymity sets,'' in \emph{Proceedings of the 2003 ACM Workshop on Privacy in
  the Electronic Society}.\hskip 1em plus 0.5em minus 0.4em\relax New York, NY,
  USA: Association for Computing Machinery, 2003, p. 72–78.

\bibitem{7870221}
A.~Belozubova, A.~Epishkina, and K.~Kogos, ``Random delays to limit timing
  covert channel,'' in \emph{2016 European Intelligence and Security
  Informatics Conference (EISIC)}, 2016, pp. 188--191.

\bibitem{bhagoji19icml}
A.~N. Bhagoji, S.~Chakraborty, P.~Mittal, and S.~Calo, ``Analyzing federated
  learning through an adversarial lens,'' in \emph{International Conference on
  Machine Learning}.\hskip 1em plus 0.5em minus 0.4em\relax ICML, 2019, pp.
  634--643.

\bibitem{secureAggregation}
K.~Bonawitz, V.~Ivanov, B.~Kreuter, A.~Marcedone, H.~B. McMahan, S.~Patel,
  D.~Ramage, A.~Segal, and K.~Seth, ``Practical secure aggregation for
  privacy-preserving machine learning,'' in \emph{Proceedings of the 2017 ACM
  SIGSAC Conference on Computer and Communications Security}, 2017, pp.
  1175--1191.

\bibitem{965114}
G.~Burel, C.~Bouder, and O.~Berder, ``Detection of direct sequence spread
  spectrum transmissions without prior knowledge,'' in \emph{GLOBECOM'01. IEEE
  Global Telecommunications Conference}, 2001.

\bibitem{burkhalter2021rofl}
L.~Burkhalter, A.~Viand, N.~K{\"u}chler, A.~Hithnawi \emph{et~al.}, ``Rofl:
  Attestable robustness for secure federated learning,'' \emph{arXiv preprint
  arXiv:2107.03311}, 2021.

\bibitem{IP_covert_timming}
S.~Cabuk, C.~E. Brodley, and C.~Shields, ``Ip covert timing channels: Design
  and detection,'' in \emph{Proceedings of the 11th ACM Conference on Computer
  and Communications Security}, ser. CCS '04.\hskip 1em plus 0.5em minus
  0.4em\relax New York, NY, USA: Association for Computing Machinery, 2004, p.
  178–187.

\bibitem{Chen2019DetectingBA}
B.~Chen, W.~Carvalho, N.~Baracaldo, H.~Ludwig, B.~Edwards, T.~Lee, I.~Molloy,
  and B.~Srivastava, ``Detecting backdoor attacks on deep neural networks by
  activation clustering,'' \emph{ArXiv}, vol. abs/1811.03728, 2019.

\bibitem{Chen2017TargetedBA}
X.~Chen, C.~Liu, B.~Li, K.~Lu, and D.~Song, ``Targeted backdoor attacks on deep
  learning systems using data poisoning,'' \emph{ArXiv}, vol. abs/1712.05526,
  2017.

\bibitem{covertChannelAttacksFL}
G.~Costa, F.~Pinelli, S.~Soderi, and G.~Tolomei, ``Covert channel attack to
  federated learning systems,'' \emph{arXiv preprint arXiv:2104.10561}, 2021.

\bibitem{Cover06}
T.~M. Cover and J.~A. Thomas, \emph{Elements of Information Theory}.\hskip 1em
  plus 0.5em minus 0.4em\relax Wiley \& Sons, 2006.

\bibitem{7087183}
D.~M. Dakhane and P.~R. Deshmukh, ``Active warden for tcp sequence number base
  covert channel,'' in \emph{2015 International Conference on Pervasive
  Computing (ICPC)}, 2015, pp. 1--5.

\bibitem{de_gaspari_evading_2022}
F.~De~Gaspari, D.~Hitaj, G.~Pagnotta, L.~De~Carli, and L.~V. Mancini, ``Evading
  behavioral classifiers: A comprehensive analysis on evading ransomware
  detection techniques,'' \emph{Neural Computing and Applications}, vol.~34,
  no.~14, pp. 12\,077--12\,096, Jul. 2022.

\bibitem{de_gaspari_reliable_2022}
------, ``Reliable detection of compressed and encrypted data,'' \emph{Neural
  Computing and Applications}, vol.~34, no.~22, pp. 20\,379--20\,393, Nov.
  2022.

\bibitem{nlp1}
J.~Devlin, M.~Chang, K.~Lee, and K.~Toutanova, ``{BERT:} pre-training of deep
  bidirectional transformers for language understanding,'' in \emph{Proceedings
  of the 2019 Conference of the North American Chapter of the Association for
  Computational Linguistics: Human Language Technologies, {NAACL-HLT} 2019},
  2019, pp. 4171--4186.

\bibitem{characterizing_network_covert_channels}
K.~Eggers and P.~Mallett, ``Characterizing network covert storage channels,''
  in \emph{Fourth Aerospace Computer Security Applications}.\hskip 1em plus
  0.5em minus 0.4em\relax Los Alamitos, CA, USA: IEEE Computer Society, 1988,
  pp. 275--279.

\bibitem{stego_active_warden}
G.~Fisk, M.~Fisk, C.~Papadopoulos, and J.~Neil, ``Eliminating steganography in
  internet traffic with active wardens,'' in \emph{Revised Papers from the 5th
  International Workshop on Information Hiding}.\hskip 1em plus 0.5em minus
  0.4em\relax Berlin, Heidelberg: Springer-Verlag, 2002, p. 18–35.

\bibitem{fredrikson2015model}
M.~Fredrikson, S.~Jha, and T.~Ristenpart, ``Model inversion attacks that
  exploit confidence information and basic countermeasures,'' in
  \emph{Proceedings of the 22nd ACM SIGSAC conference on computer and
  communications security}, 2015, pp. 1322--1333.

\bibitem{cnn_citation}
K.~Fukushima, ``{Neocognitron: A self-organizing neural network model for a
  mechanism of pattern recognition unaffected by shift in position},''
  \emph{Biological Cybernetics}, 1980.

\bibitem{ganju2018property}
K.~Ganju, Q.~Wang, W.~Yang, C.~A. Gunter, and N.~Borisov, ``Property inference
  attacks on fully connected neural networks using permutation invariant
  representations,'' in \emph{Proceedings of the 2018 ACM SIGSAC Conference on
  Computer and Communications Security}, 2018, pp. 619--633.

\bibitem{Gianvecchio}
S.~Gianvecchio, H.~Wang, D.~Wijesekera, and S.~Jajodia, ``Model-based covert
  timing channels: Automated modeling and evasion.''

\bibitem{griffin_tcp_timestamps}
J.~Giffin, R.~Greenstadt, P.~Litwack, and R.~Tibbetts, ``Covert messaging
  through tcp timestamps,'' in \emph{Proceedings of the 2nd International
  Conference on Privacy Enhancing Technologies}, ser. PET'02.\hskip 1em plus
  0.5em minus 0.4em\relax Berlin, Heidelberg: Springer-Verlag, 2002, p.
  194–208.

\bibitem{covert_channels}
V.~Gligor, \emph{Covert Channel Analysis of Trusted Systems. A Guide to
  Understanding}.\hskip 1em plus 0.5em minus 0.4em\relax National Computer
  Security Center, 1993.

\bibitem{norm_computation}
G.~H. Golub and C.~F. Van~Loan, \emph{Matrix Computations, 4th Edition}.\hskip
  1em plus 0.5em minus 0.4em\relax John Hopkins University Press, 2013.

\bibitem{Goodfellow_deeplearning_book}
I.~Goodfellow, Y.~Bengio, and A.~Courville, \emph{Deep Learning}.\hskip 1em
  plus 0.5em minus 0.4em\relax MIT Press, 2016,
  \url{http://www.deeplearningbook.org}.

\bibitem{speech2}
A.~Graves, A.~Mohamed, and G.~Hinton, ``Speech recognition with deep recurrent
  neural networks,'' in \emph{2013 IEEE International Conference on Acoustics,
  Speech and Signal Processing}, 2013, pp. 6645--6649.

\bibitem{Gu2017BadNetsIV}
T.~Gu, K.~Liu, B.~Dolan-Gavitt, and S.~Garg, ``Badnets: Evaluating backdooring
  attacks on deep neural networks,'' \emph{IEEE Access}, vol.~7, pp.
  47\,230--47\,244, 2019.

\bibitem{He2016DeepRL}
K.~He, X.~Zhang, S.~Ren, and J.~Sun, ``Deep residual learning for image
  recognition,'' \emph{2016 IEEE Conference on Computer Vision and Pattern
  Recognition (CVPR)}, pp. 770--778, 2016.

\bibitem{speech3}
G.~Hinton, L.~Deng, D.~Yu, G.~E. Dahl, A.-r. Mohamed, N.~Jaitly, A.~Senior,
  V.~Vanhoucke, P.~Nguyen, T.~N. Sainath, and B.~Kingsbury, ``Deep neural
  networks for acoustic modeling in speech recognition: The shared views of
  four research groups,'' \emph{IEEE Signal processing magazine}, vol.~29,
  no.~6, pp. 82--97, 2012.

\bibitem{hitaj2017deep}
B.~Hitaj, G.~Ateniese, and F.~P{\'e}rez-Cruz, ``Deep models under the {GAN}:
  Information leakage from collaborative deep learning,'' in \emph{Proceedings
  of the 2017 ACM SIGSAC Conference on Computer and Communications
  Security}.\hskip 1em plus 0.5em minus 0.4em\relax ACM, 2017, pp. 603--618.

\bibitem{passGAN}
B.~Hitaj, P.~Gasti, G.~Ateniese, and F.~Perez-Cruz, ``Passgan: A deep learning
  approach for password guessing,'' in \emph{Applied Cryptography and Network
  Security}, R.~H. Deng, V.~Gauthier-Uma{\~{n}}a, M.~Ochoa, and M.~Yung,
  Eds.\hskip 1em plus 0.5em minus 0.4em\relax Cham: Springer International
  Publishing, 2019, pp. 217--237.

\bibitem{9257172}
D.~Hitaj, B.~Hitaj, S.~Jajodia, and L.~V. Mancini, ``Capture the bot: Using
  adversarial examples to improve captcha robustness to bot attacks,''
  \emph{IEEE Intelligent Systems}, vol.~36, no.~5, pp. 104--112, 2021.

\bibitem{hitaj2023minerva}
D.~Hitaj, G.~Pagnotta, F.~D. Gaspari, L.~D. Carli, and L.~V. Mancini,
  ``Minerva: A file-based ransomware detector,'' 2023.

\bibitem{lstm_citation}
S.~Hochreiter and J.~Schmidhuber, ``{Long Short-Term Memory},'' \emph{Neural
  Computation}, vol.~9, no.~8, pp. 1735--1780, 1997.

\bibitem{jere2020taxonomy}
M.~S. Jere, T.~Farnan, and F.~Koushanfar, ``A taxonomy of attacks on federated
  learning,'' \emph{IEEE Security \& Privacy}, vol.~19, no.~2, pp. 20--28,
  2020.

\bibitem{kairouz2019advances}
P.~Kairouz, H.~B. McMahan, B.~Avent, A.~Bellet, M.~Bennis, A.~N. Bhagoji,
  K.~Bonawitz, Z.~Charles, G.~Cormode, R.~Cummings \emph{et~al.}, ``Advances
  and open problems in federated learning,'' \emph{arXiv preprint
  arXiv:1912.04977}, 2019.

\bibitem{karger}
P.~A. Karger and J.~C. Wray, ``Storage channels in disk arm optimization,'' in
  \emph{IEEE Symposium on Security and Privacy}.\hskip 1em plus 0.5em minus
  0.4em\relax IEEE Computer Society, 1991.

\bibitem{styleGAN}
T.~Karras, S.~Laine, and T.~Aila, ``A style-based generator architecture for
  generative adversarial networks,'' in \emph{2019 IEEE/CVF Conference on
  Computer Vision and Pattern Recognition (CVPR)}, 2019, pp. 4396--4405.

\bibitem{krizhevsky2009learning}
A.~{Krizhevsky}, ``Learning multiple layers of features from tiny images,''
  \emph{Technical Report TR-2009, University of Toronto}, 2009.

\bibitem{lampson_confinement}
B.~W. Lampson, ``A note on the confinement problem,'' \emph{Commun. ACM},
  vol.~16, p. 613–615, 1973.

\bibitem{mnist_dataset}
Y.~LeCun and C.~Cortes, ``{MNIST} handwritten digit database,''
  \url{http://yann.lecun.com/exdb/mnist/}, 2010.

\bibitem{covert_channel_ad_hoc}
S.~Li and A.~Epliremides, ``A network layer covert channel in ad-hoc wireless
  networks,'' in \emph{2004 First Annual IEEE Communications Society Conference
  on Sensor and Ad Hoc Communications and Networks, 2004. IEEE SECON 2004.},
  2004, pp. 88--96.

\bibitem{Liu2018FinePruningDA}
K.~Liu, B.~Dolan-Gavitt, and S.~Garg, ``Fine-pruning: Defending against
  backdooring attacks on deep neural networks,'' in \emph{International
  Symposium on Research in Attacks, Intrusions, and Defenses}.\hskip 1em plus
  0.5em minus 0.4em\relax Springer, 2018, pp. 273--294.

\bibitem{5198826}
X.~Luo, E.~W.~W. Chan, and R.~K.~C. Chang, ``Clack: A network covert channel
  based on partial acknowledgment encoding,'' in \emph{2009 IEEE International
  Conference on Communications}, 2009, pp. 1--5.

\bibitem{4630112}
------, ``Tcp covert timing channels: Design and detection,'' in \emph{2008
  IEEE International Conference on Dependable Systems and Networks With FTCS
  and DCC (DSN)}, 2008, pp. 420--429.

\bibitem{luping2019cmfl}
W.~Luping, W.~Wei, and L.~Bo, ``{CMFL}: Mitigating communication overhead for
  federated learning,'' in \emph{2019 IEEE 39th International Conference on
  Distributed Computing Systems (ICDCS)}.\hskip 1em plus 0.5em minus
  0.4em\relax IEEE, 2019, pp. 954--964.

\bibitem{lyu2020threats}
L.~Lyu, H.~Yu, J.~Zhao, and Q.~Yang, ``Threats to federated learning,'' in
  \emph{Federated Learning}.\hskip 1em plus 0.5em minus 0.4em\relax Springer,
  2020, pp. 3--16.

\bibitem{valenciacovidwired2021}
W.~Marx, ``How valencia crushed covid with ai,''
  \url{https://www.wired.co.uk/article/valencia-ai-covid-data}, 2021.

\bibitem{mcmahan2017communication}
B.~McMahan, E.~Moore, D.~Ramage, S.~Hampson, and B.~A. y~Arcas,
  ``Communication-efficient learning of deep networks from decentralized
  data,'' in \emph{Artificial Intelligence and Statistics}.\hskip 1em plus
  0.5em minus 0.4em\relax PMLR, 2017, pp. 1273--1282.

\bibitem{mcmahangboard2017}
B.~McMahan and D.~Ramage, ``Federated learning: Collaborative machine learning
  without centralized training data,''
  \url{https://ai.googleblog.com/2017/04/federated-learning-collaborative.html},
  2017.

\bibitem{melis2019exploiting}
L.~Melis, C.~Song, E.~De~Cristofaro, and V.~Shmatikov, ``Exploiting unintended
  feature leakage in collaborative learning,'' in \emph{2019 IEEE Symposium on
  Security and Privacy (SP)}.\hskip 1em plus 0.5em minus 0.4em\relax IEEE,
  2019, pp. 691--706.

\bibitem{menick2018generating}
J.~Menick and N.~Kalchbrenner, ``Generating high fidelity images with subscale
  pixel networks and multidimensional upscaling,'' in \emph{International
  Conference on Learning Representations}, 2019.

\bibitem{wikitext_dataset}
S.~Merity, C.~Xiong, J.~Bradbury, and R.~Socher, ``Pointer sentinel mixture
  models,'' \emph{CoRR}, vol. abs/1609.07843, 2016.

\bibitem{machinelearningMitchell}
T.~M. Mitchell, \emph{Machine Learning}, 1st~ed.\hskip 1em plus 0.5em minus
  0.4em\relax New York, NY, USA: McGraw-Hill, Inc., 1997.

\bibitem{mo2021ppfl}
F.~Mo, H.~Haddadi, K.~Katevas, E.~Marin, D.~Perino, and N.~Kourtellis, ``Ppfl:
  Privacy-preserving federated learning with trusted execution environments,''
  \emph{arXiv preprint arXiv:2104.14380}, 2021.

\bibitem{9833616}
G.~Pagnotta, D.~Hitaj, F.~D. Gaspari, and L.~V. Mancini, ``Passflow: Guessing
  passwords with generative flows,'' in \emph{2022 52nd Annual IEEE/IFIP
  International Conference on Dependable Systems and Networks (DSN)}.\hskip 1em
  plus 0.5em minus 0.4em\relax IEEE Computer Society, jun 2022, pp. 251--262.

\bibitem{pagnotta2023dolos}
G.~Pagnotta, F.~D. Gaspari, D.~Hitaj, M.~Andreolini, M.~Colajanni, and L.~V.
  Mancini, ``Dolos: A novel architecture for moving target defense,'' 2023.

\bibitem{gallager_book}
J.~Patrick R.~Gallagher, \emph{A guide to understanding covert channel analysis
  of trusted systems}.\hskip 1em plus 0.5em minus 0.4em\relax National Computer
  Security Center, 1993, \url{https://irp.fas.org/nsa/rainbow/tg030.htm}.

\bibitem{1624024}
P.~Peng, P.~Ning, and D.~Reeves, ``On the secrecy of timing-based active
  watermarking trace-back techniques,'' in \emph{2006 IEEE Symposium on
  Security and Privacy (S P'06)}, 2006, pp. 15 pp.--349.

\bibitem{piczak2015dataset}
K.~J. Piczak, ``{ESC}: {Dataset} for {Environmental Sound Classification},'' in
  \emph{Proceedings of the 23rd {Annual ACM Conference} on {Multimedia}}.\hskip
  1em plus 0.5em minus 0.4em\relax {ACM}, 2015, pp. 1015--1018.

\bibitem{nlp2}
A.~Radford, J.~Wu, R.~Child, D.~Luan, D.~Amodei, and I.~Sutskever, ``Language
  models are unsupervised multitask learners,'' 2019.

\bibitem{Richardson08}
T.~Richardson and R.~Urbanke, \emph{Modern Coding Theory}.\hskip 1em plus 0.5em
  minus 0.4em\relax Cambridge University Press, 2008.

\bibitem{Schafer}
M.~Schaefer, B.~Gold, R.~Linde, and J.~Scheid, ``Program confinement in
  kvm/370,'' in \emph{Proceedings of the 1977 Annual Conference}, ser. ACM
  '77.\hskip 1em plus 0.5em minus 0.4em\relax Association for Computing
  Machinery, 1977, p. 404–410.

\bibitem{shokri2015privacy}
R.~Shokri and V.~Shmatikov, ``Privacy-preserving deep learning,'' in
  \emph{Proceedings of the 22nd ACM SIGSAC conference on computer and
  communications security}.\hskip 1em plus 0.5em minus 0.4em\relax ACM, 2015,
  pp. 1310--1321.

\bibitem{Simonyan14verydeep}
K.~Simonyan and A.~Zisserman, ``Very deep convolutional networks for
  large-scale image recognition,'' 2014.

\bibitem{song2017machine}
C.~Song, T.~Ristenpart, and V.~Shmatikov, ``Machine learning models that
  remember too much,'' in \emph{Proceedings of the 2017 ACM SIGSAC Conference
  on computer and communications security}, 2017, pp. 587--601.

\bibitem{spread_spectrum_principles}
D.~Torrieri, \emph{Principles of Spread-Spectrum Communication Systems, 4th
  Edition}.\hskip 1em plus 0.5em minus 0.4em\relax Springer, Cham, 2018.

\bibitem{Tran2018SpectralSI}
B.~Tran, J.~Li, and A.~Madry, ``Spectral signatures in backdoor attacks,'' in
  \emph{NeurIPS}, 2018.

\bibitem{triastcyn2019federated}
A.~Triastcyn and B.~Faltings, ``Federated learning with bayesian differential
  privacy,'' in \emph{2019 IEEE International Conference on Big Data (Big
  Data)}.\hskip 1em plus 0.5em minus 0.4em\relax IEEE, 2019, pp. 2587--2596.

\bibitem{Verdu98}
S.~Verdu, \emph{Multiuser Detection}.\hskip 1em plus 0.5em minus 0.4em\relax
  Cambridge University Press, 1998.

\bibitem{wang2020attack}
H.~Wang, K.~Sreenivasan, S.~Rajput, H.~Vishwakarma, S.~Agarwal, J.-y. Sohn,
  K.~Lee, and D.~Papailiopoulos, ``Attack of the tails: Yes, you really can
  backdoor federated learning,'' \emph{Advances in Neural Information
  Processing Systems}, 2020.

\bibitem{netwarden}
J.~Xing, Q.~Kang, and A.~Chen, ``{NetWarden}: Mitigating network covert
  channels while preserving performance,'' in \emph{29th USENIX Security
  Symposium (USENIX Security 20)}.\hskip 1em plus 0.5em minus 0.4em\relax
  USENIX Association, Aug. 2020, pp. 2039--2056.

\bibitem{zhang2019gradient}
J.~Zhang, T.~He, S.~Sra, and A.~Jadbabaie, ``Why gradient clipping accelerates
  training: A theoretical justification for adaptivity,'' in
  \emph{International Conference on Learning Representations}, 2019.

\bibitem{zhang2021leakage}
W.~Zhang, S.~Tople, and O.~Ohrimenko, ``Leakage of dataset properties in
  multi-party machine learning,'' in \emph{30th {USENIX} Security Symposium
  ({USENIX} Security 21)}.\hskip 1em plus 0.5em minus 0.4em\relax {USENIX}
  Association, 2021, pp. 2687--2704.

\end{thebibliography}
